\documentclass[a4paper,fleqn,usenatbib]{mnras}

\usepackage[T1]{fontenc}
\usepackage{ae,aecompl}

\usepackage{savesym}
\usepackage{graphicx}	
\usepackage{amsmath}	
\savesymbol{iint}
\usepackage{amssymb}
\usepackage{txfonts}
\restoresymbol{TXF}{iint}

\usepackage{multicol}        
\usepackage{bm}	
\usepackage{rotating}
\title{On the scaling relations of bulges and early type galaxies}
\author[B. A. Pastrav]{Bogdan A. Pastrav$^{1}$\thanks{E-mail:bapastrav@spacescience.ro}
\\
$^{1}$Cosmology and Astroparticle Physics Laboratory, Institute of Space Science, Atomistilor 409, 077125, Bucharest-Magurele, Romania\\
}
\date{Accepted 2021 June 15. Received 2021 June 11; in original form 2020 December 21.}

\pubyear{2020}

\begin{document}
\label{firstpage}
\pagerange{\pageref{firstpage}--\pageref{lastpage}}
\maketitle

\begin{abstract}
Following from our recent work, we present here a detailed structural analysis of a representative sample of nearby spiral and early-type galaxies taken from the KINGFISH/
SINGS survey. The photometric parameters of bulges are obtained from bulge-disc decompositions using GALFIT data analysis algorithm. The method and corrections for projection
and dust effects previously obtained are used to derive intrinsic photometric and structural bulge parameters. We show the main bulge scaling relations and the black hole 
relations, both observed and intrinsic ones, in B band. We find dust and inclination effects to produce more important changes in the parameters of the Kormendy
relation for spiral galaxies, with the respective bulges of late-type galaxies residing on a steeper slope relation that the early-type galaxies. We observe
that the Kormendy relation in combination with a bulge S\'{e}rsic index ($n_{b}$) threshold, does not produce a conclusive morphological separation of bulges. The
$n_{b}$- bulge-to-total flux ratio ($B/T$) and $B/T$-stellar mass could be used to discriminate between late-type and early-type galaxies, while a further use
of these parameters to divide bulges with different morphologies is problematic due to overlaps in the two distributions or large spread in values. We confirm the existence of
two distinct intrinsic relations between the bulge luminosity (or absolute magnitude) and S\'{e}rsic index for late- and early-type galaxies, while the relations between
the black-hole mass ($M_{BH}$) and bulge luminosity are not found to be statistically different at p<0.05. Within errors, we find statistically similar intrinsic $M_{BH}-n_{b}$ 
relations for all bulges. 

\end{abstract}

\begin{keywords}
  galaxies: bulges -- galaxies: elliptical and lenticular, cD -- galaxies: structure -- ISM: dust, extinction -- galaxies: evolution -- galaxies: fundamental parameters
 \end{keywords}

 \maketitle
 
\section{Introduction}
 
Bulges, traditionally defined as the excess of light above the inner extrapolation of discs and/or bars, were for a long time thought to be elliptical-like structures, 
predominantly pressure-supported, embedded in the galaxy disc. These components - the classical bulges - have similar observational properties as elliptical galaxies (Faber 1977,
Gott 1977, Renzini 1999) and form out of major mergers (Brooks \& Christensen 2016, Tonini et al. 2016, Rodriguez-Gomez et al. 2017) or coalescence of giant clumps (Bournaud et al.
2007, Bournaud et al. 2009, Elmegreen et al. 2008, Bournaud et al. 2016). However, many studies have shown the existence of a dichotomy of bulges (Kormendy 2003, Kormendy
\& Kennicutt 2004, Athanassoula 2005, Fisher \& Drory 2016), with the second category of bulges - named pseudo-bulges, being flat, disc-like stuctures, rotation-supported and
with younger stellar populations than classical bulges. It was also observed a coexistence of the two types of bulges within the same galaxy (Fisher \& Drory 2008,  Erwin et al. 2015).\\
Understanding the formation scenarios of bulges and their subsequent evolution is of critical importance in the galaxy formation and evolution models. The scaling relations of 
these components are also essential. Thus, the classical bulges follow the same scaling relations as ellipticals, such as Faber-Jackson (Faber \& Jackson 1976) relation, the
Fundamental Plane (Djorkovsky \& Davis 1987) and its projection - the Kormendy (Kormendy 1977) relation. It has also been shown the connection of bulges with the central
massive black holes from the centre of galaxies, their masses being correlated (Gebhardt et al.  2000, Ferrarese \& Merritt 2000, Kormendy \& Gebhardt 2001). The associated
scaling relations, the black-hole scaling relations, are also very important for the understanding of the co-evolution of galaxies and the central black holes.

Some of these relations and/or bulge structural and photometric parameters were used alone or in combination to separate bulges of smaller or larger samples of galaxies into
classical or pseudo-bulges. Thus, \cite{Fish08} used the bulge S\'{e}rsic index ($n_{b}$) for this division, while \cite{Gad09} and \cite{Gao20} argued that a classification
using this parameter alone, produces an important degree of contamination in the two categories of bulges. \cite{All06} used a combination of bulge-to-total flux ratios ($B/T$)
and the ratio between bulge and disc sizes. \cite{Neu17} on the other hand, used a combination of four parameters to classify bulges and found the Kormendy relation together
with their custom concentration index to provide the most accurate classification. \cite{Gad09}, \cite{Sach17}, \cite{Sach19} and \cite{Gao20} used the Kormendy relation of the
eliptical galaxies to classify bulges and found with a high degree of accuracy that pseudo-bulges are the ones which are situated far (more than two or three standard
deviations) from the aforementioned relation. \cite{Sach20} defined a new morphology indicator from the Kormendy relation - the relative distance from the $\pm1\sigma$ boundary 
of the Kormendy relation, and used it to investigate the degree of accuracy of other bulge morphology classifiers (e.g. $n_{b}$, $B/T$, concentration index or central velocity 
dispersion). They found this to be the most efficient indicator of bulge morphology and stellar activity. Despite all this, as pointed out by \cite{Neu17}, there is not a
single criterion that should be used when trying to distinguish between classical and pseudo-bulges, but a combination of them for a more robust classification, as one type
of bulges may not follow all the criteria or have all the morphological/structural properties usually characteristic for them (see Fisher \& Drory 2016, Kormendy 2016).\\
Added to all this, it has been shown by \cite{Tuf04}, \cite{Mol06}, \cite{Gad10}, \cite{Pas13a} and \cite{Pas13b} that dust (which pervades the disc) and projection (inclination) 
effects produce biases in the parameters involved in the scaling relations of bulges and early-type galaxies, or in those employed when trying to separate bulges with different 
morphologies. These biases are more sigmificant at shorter wavelengths and higher inclinations. Not taking into account these effects may produce biased scaling relations, 
leading to misleading conclusions, misclassifications of bulges, or validation/invalidation of certain criteria for bulge morphology classification.

Thus, in this paper we continue the study presented in \cite{Pas20} (hereafter, Paper I) and show the results of a detailed analysis of a sample of low-redhsift spiral and
early-type galaxies taken from the KINGFISH survey (Kennicutt et al. 2011). Here we focus on the scaling relations of bulges of spirals and early-type galaxies, showing how
dust, inclination and decomposition effects bias the specific parameters of the relations, such as slope, zero point or scatter. Special attention is dedicated to a set of scaling
relations, important for understanding the connection between bulges and black holes. For certain relations, we investigate the degree of correlation between the parameters 
involved. We also investigate, in relation with previous works, which relations or combination of parameters can produce a more accurate classification of galaxy bulges into 
classical or pseudo-bulges, even though our sample is rather small. We then compare our main results with other similar works in the literature. As before, we decompose each galaxy
into its main components (bulge+disc). Then, we use the method of \cite{Pas13a} and \cite{Pas13b} and their numerical corrections for projection (inclination), dust and 
decomposition effects, to recover the intrinsic photometric and structural paraneters involved in the scaling relations. The numerical corrections were derived by analysing
and fitting simulated images of galaxies produced by means of radiative transfer calculations and the model of \cite{Pop11}. The empirical relation found by \cite{Gro13} is
used to determine the central face-on dust opacity, a parameter which is essential when applying the corrections for dust effects, in the case of bulges and lenticular galaxies.

The paper is organised as follows. In Sect.~\ref{sec:sample} we present the galaxy sample used in this study, while in Sect.~\ref{sec:method} we describe the method used for 
the overall procedure (photometry, fitting algorithm, derive dust opacity and dust mass, derive intrinsic parameteres needed in the scaling relations). In Sect.~\ref{sec:results}
we present the main results - the galaxy scaling relations for bulges of spiral galaxies and the early-type ones, both observed and intrinsic (dust-free) ones, together with
all the numerical results.  The black-hole scaling relations are presented and analysed in Sect.~\ref{sec:black-hole}. In Sect.~\ref{sec:discussion} we discuss 
upon the main results and possible sources of errors and differences with other studies, while in Sect.~\ref{sec:conclusions} we summarise the results obtained in 
this study and draw conclusions.\\
Throughout this paper, where necessary, a Hubble constant of $H_{0}=67.8$ km/s/Mpc (Planck Collaboration 2016) was used.

\section{Sample}\label{sec:sample}
 
Our sample consists of the original sample of 18 late-type galaxies analysed in Paper I, to which another 11 early-type galaxies (6 ellipticals and 5 lenticulars) were added,
the latter also included in the SINGS (\textit{Spitzer} Infrared Nearby Galaxies Survey; Kennicutt et al. 2003) survey and the KINGFISH project (Key Insights on Nearby 
Galaxies: a Far-Infrared Survey with \textit{Herschel}; Kennicutt et al. 2011). Thus, in total our sample consists of 29 galaxies, which is used to study the scaling relations 
of both bulges and early-type galaxies at low redshift.
The KINGFISH project is an imaging and spectroscopic survey, consisting of 61 nearby (d<30 Mpc) galaxies, chosen to cover a wide range of galaxy properties (morphologies, 
luminosities, SFR, etc.) and local ISM environments characteristic for the nearby universe. The optical images were extracted from the NASA/IPAC Infrared Science Archive 
(IRSA) and NASA IPAC Extragalactic Database (NED). The images were taken with the KPNO (Kitt Peak National Observatory) and CTIO (Cerro Tololo Inter-American Observatory) 
telescopes (see Kennicutt et al. (2003)). \\
As in \cite{Pas20}, barred galaxies present in the KINGFISH survey were excluded from our selected samples because at the moment we cannot properly account and correct for 
the effects of dust on the photometric and structural parameters of bars. The images analysed here are in optical regime, in B band, for a number of reasons.
Thus, the galaxy images in the short optical bands (e.g B, r) present a much higher resolution, and are better resolved overall and up to larger galactocentric radii than
at longer wavelengths. Thus, on one hand, one can analyse and disentangle the galaxy components with a higher degree of accuracy. On the other hand, as short optical wavelength images
can be obtained up to a considerably deeper (lower) surface brightness limit, one can go deep into the backround when analysing images at these wavelengths and have a more correct
picture of the galaxy outskirts and possible truncations in the galaxy surface brightness profiles at large galactocentric radii.
Moreover, as dust and inclination effects are stronger at shorter wavelengths (as shown in Pastrav et al. (2013a), Pastrav et al. (2013b)), and because our method is suitable 
for cases where optical data are available, we have chosen the B band images.

\section{Method}\label{sec:method}

The method used in this study is in general the same as in Paper I when it comes to the fitting procedure, the sky determination and subtraction, the photometry and the 
dust opacity and dust mass derivation, with a few differences. Therefore, for a more detailed description, we refer the reader to Paper I, where the whole procedure is presented 
in great detail. Here, a shorter and concise version is given below.

\subsection{Fitting procedure}\label{sec:fitting}

For the fitting procedure of the early-type galaxies (ETGs) we used the GALFIT (version 3.0.2) data analysis algorithm (Peng et al. 2002, Peng et al. 2010), as we
did for the case of spirals in our sample. The analysis of the bulges of spiral galaxies was done in Paper I when analysing the sample of late-type galaxies. 
Thus, in this study we just analysed the ellipticals and lenticulars added here for the purpose of studying their scaling relations. For the bulge-to-disc decomposition 
and to fit the observed surface brightness of ETGs we used the exponential (``expdisc'') and the  S\'{e}rsic (``sersic'') functions available in GALFIT, for the disc and bulge 
surface brightness profiles, while the "sky" function was used for an estimation of the background in each image.

In most studies from the literature, elliptical galaxies are usually considered as a single component, a spheroid, and thus fitted with a single S\'{e}rsic function
only. This approach may be accurate, especially at lower resolution and for higher redshift galaxy images. Hovewer, it has also been shown in quite a few studies, such as
the ones of \cite{Capa87}, \cite{Ben90}, \cite{Rix90}, \cite{Rix92} or \cite{Nie91} that ellipticals can present a stellar disc in their structure or intermediate-scale discs.
\cite{Hua13a}, \cite{Hua13b}, \cite{Hua16} found the structure of local, bright elliptical galaxies to be formed by three sub-components, with the surface brightness of each being
described by a S\'{e}rsic function. The works of \cite{Bell17}, \cite{Oh17} and \cite{Zhu21} came to emphasize that ellipticals are not just simple one-component structures and 
the need to consider extra-components in their structural analysis when trying to explain their evolutionary history. \cite{Oh17} found that using a single-component to model
the surface brightness of the ellipticals cannot always match the ellipticity curve, or alternatively, changes in the isophotal shape.
Thus, in this study, as our images are well resolved and at high resolution and signal-to-noise ratio, we decided to use the same combination of functions as for spirals and
lenticulars.

As in our previous work, the free parameters of the fits are: the X and Y coordinates of the centre of the galaxy in pixels, the bulge and disc integrated magnitudes, the 
disc scale-length / bulge effective radius (for exponential/S\'{e}rsic function), axis-ratios of discs and bulges, bulge S\'{e}rsic index (for S\'{e}rsic function), the sky background
(only in the preliminary fit - Step 1, see Paper I) and the sky gradients in X and Y. The input values for the coordinates of galaxy centre were determined after a careful inspection 
of each image. Initial values for the position angles (PA) and axis-ratios were taken from NED. Although the central coordinates are free parameters, we imposed a constraint
on the fitting procedure, ensuring that the bulge and disc components were centred on the same position. The axis-ratio is defined as the ratio between the semi-minor and semi-major axis of
the model fit (for each component). The position angle is the angle between the semi-major axis and the Y axis (increasing counter clock-wise). To mask the pixels corresponding to the
additional light coming from neighboring galaxies, stars, compact sources, AGN or image artifacts, for each galaxy image we used a complex star-masking routine to create a bad pixel mask.
This was used as input in GALFIT.

\subsection{Sky determination and subtraction. Photometry}\label{sec:sky}

We follow the procedure with the steps described in Paper I (Step 1, 2 and, where necessary - Step 3) to estimate the background level, derive the structural and photometric parameters, and
calculate the integrated fluxes for each galaxy, together with the bulge-to-disc ratios. The integrated (total) flux of the galaxy is calculated from the maximum
curve-of-growth (CoG) value (in counts), at the $R_{max}$ radius (this is defined as the radius beyond which there is no galaxy emission and, therefore, the CoG is essentially 
flat towards larger radii). The bulge-to-disc ratio ($B/D$) is estimated from the disc and bulge CoGs and compared with the one determined by the ratio of the total counts of the decomposed 
disc and bulge images, as it has to be consistent, within errors. The uncertainties in the fluxes are estimated from the root mean square of the CoG values from the first 
10 elliptical anulli beyond $R_{max}$. As before, we have used the positive sky residuals in the outer parts of galaxies (towards $R_{max}$ and beyond) to estimate the systematic 
errors in bulge-to-disc ratios. This is important because the sky level errors \text{have an important contribution to} the systematic errors in bulge-to-disc decompositions, 
as shown by \cite{Sim02}. In addition, in nearby, well resolved, high resolution galaxy images, the choice of modelling functions for the surface distribution can
produce a more significant contribution to the error budget and the best-fit bulge parameters (including the bulge-to-total flux ratio), like the studies of \cite{Lau05}, 
\cite{Kim14} and \cite{Gao17} have determined. \\
The derived integrated fluxes and bulge-to-disc ratios for the ETGs of our sample are given in Table~\ref{tab:photo_fluxes}, together with the distances to each galaxy used
in this study, taken from NED.

With the best-fitting structural/photometric parameters and the integrated fluxes for the galaxies and their main constituents already determined, we then calculated the central, effective
and mean effective surface brightness for bulges, together with the corresponding apparent and absolute bulge magnitudes. For the bulges of the spiral galaxies, these were already 
calculated in our previous paper (see Table 4 in Pastrav (2020)). Thus, following for example \cite{Gra05} or \cite{Gra08}, we can write the equations for the bulge effective 
surface brightness, central and mean effective surface brightness, bulge apparent and absolute magnitudes as:
\begin{eqnarray}\label{eq:SBe_b}
 \mu_{e,b}=-2.5\log[\frac{F_{b}}{2\pi(R_{e,b})^2\exp(\kappa_{n})n\kappa_{n}^{-2n}\Gamma(2n)Q_{b}}/F_{0}] 
\end{eqnarray}
\begin{eqnarray}\label{eq:SB0_b}
 \mu_{0,b}=\mu_{e,b}-2.5\times(\kappa_{n})/log(10)
\end{eqnarray}
\begin{eqnarray}\label{eq:avg_SBe_b}
 <\mu_{e,b}>=\mu_{e,b}-2.5\log([\frac{n\exp(\kappa_{n})\Gamma(2n)}{\kappa_{n}^{2n}}]
\end{eqnarray}
where $F_{b}$ is the integrated flux of the bulge, $R_{e,b}$ is the effective radius (in arcsecs), $\Gamma(2n)=2\gamma(2n,\kappa_{n})$ (Ciotti 1991), with $\Gamma$ and $\gamma$ the 
complete and incomplete gamma functions, $n$ is the S\'{e}rsic 
index, while $\kappa_{n}$ is a variable coupled with $n$ (see Ciotti \& Bertin 1999 and Graham \& Driver 2005)
\begin{eqnarray}\label{eq:mapp_b}
 m_{b}=\mu_{e,b}-2.5\log(2\pi(R_{e,b})^2Q_{b})-2.5\log[\frac{n\exp(\kappa_{n})\Gamma(2n)}{\kappa_{n}^{2n}}] 
\end{eqnarray}
\begin{eqnarray}\label{eq:Mabs_b}
 M_{b}=m_{b}-25-5\log(d_{gal}/Mpc)
\end{eqnarray}
The observed axis-ratio of the bulge ($Q_{b}$) is included in Eqs.~\ref{eq:SB0_b}\&\ref{eq:mapp_b} because bulges are seen in projection. Thus, this has to be taken into
consideration later when deriving the intrinsic parameters, and correct for projection effects.\\
\begin{table}
\caption{\label{tab:photo_fluxes} The calculated fluxes for the early-type galaxies (B band). The columns represent: (1) - galaxy name; (2) - morphological type; 
(3) - distance to each galaxy, taken from NASA Extragalactic Database (NED), as derived in: $a$ - \protect\cite{Tul13}, $b$ - \protect\cite{Ken11}, $c$ - \protect\cite{Bla01}, $d$ - 
\protect\cite{Theu07}, $e$ - \protect\cite{Tul88}, $f$ - \protect\cite{Sabbi18}; (4) bulge-to-disc ratios ($B/D$) derived from the decomposed images, with systematic 
uncertainties, derived as described in Sec.~\ref{sec:sky}; (5) - the integrated flux for each galaxy, in $Jy$; (6) - the error for the galaxy flux; (7) - 
the integrated flux of the bulge, in $Jy$.}
 \begin{tabular}{{r|r|r|r|r|r|r}}
  \hline \hline
    $Galaxy$   &  $Type$   &  $d_{gal}$  &  $B/D$    &   $F_{gal}$  &   $\sigma_{F_{gal}}$   & $F_{b}$\\  
               &           &   [Mpc]   &                &      [Jy]           &        [Jy]                         &      [Jy]     \\  
   \hline 
NGC0584  &  E4 &     $20.00^{a}$  &	  $2.02_{-0.00}^{+0.00}$    &              0.310    &     0.006        &       0.190\\  
NGC0855  &  E  &     $8.83^{a}$  &       $0.37_{-0.00}^{+0.00}$     &              0.025    &     0.001        &       0.007\\   
NGC1404  &  E1(2) &  $19.50^{a}$  &       $5.52_{-0.03}^{+0.05}$    &        	   0.630       &  0.004        &       0.530\\  
NGC3265  &  E	  &  $19.60^{b}$  &       $0.72_{-0.00}^{+0.00}$   &              0.025      &   0.009        &       0.013\\ 
NGC4125  &  E6pec &  $23.90^{c,a}$  &       $3.40_{-0.05}^{+0.06}$    &              0.300      &   0.003        &       0.230\\  
NGC4552  &  E0-1  &  $15.90^{a}$  &       $0.93_{-0.08}^{+0.02}$    &              0.400      &   0.003        &       0.190\\  
NGC1377  &  S0    &  $21.00^{d}$  &       $1.63_{-0.01}^{+0.02}$    &              0.023      &   0.001        &       0.014\\  
NGC1482  &  SA0(p)& $19.60^{e}$  &       $1.11_{-0.00}^{+0.00}$    &              0.172     &    0.002    &           0.094\\   
NGC1705  &  SA0p  & $5.22^{f,a}$  &       $1.58_{-0.00}^{+0.00}$     &              0.105      &   0.004       &        0.064\\  
NGC3773  &  SA0   & $17.00^{e}$  &       $0.54_{-0.08}^{+0.08}$     &              0.030      &   0.001       &        0.011\\  
NGC5866  &  S0    & $14.70^{a}$  &       $0.36_{-0.05}^{+0.06}$     &              0.231      &   0.001       &        0.060\\   
\hline
 \end{tabular}
\end{table}


\subsection{Correcting for dust, projection and decomposition effects}\label{sec:corr}

In Paper I (see section 3.3), we showed the procedure and equations used to derive the dust opacity and dust mass for the analysed sample of spiral galaxies. 
Thus, we have used the correlation between the central face-on B band dust opacity, $\tau_{B}^{f}$, and the stellar mass surface density of nearby galaxies ($\mu_{*}$, 
derived using the scale-length found in the bulge-disc decomposition), found by \cite{Gro13}. This relation was obtained by analysing a sample of spiral galaxies taken from
the Galaxy and Mass Assembly (GAMA) survey. 
\begin{eqnarray}\label{eq:Grootes}
\log(\tau_{B}^{f})=1.12(\pm0.11)\cdot\log(\mu_{*}/M_{\odot}kpc^{-2})-8.6(\pm0.8)
\end{eqnarray}
The dust masses for each galaxy were derived using Eq.~(2) from \cite{Gro13} (see also Eqs.~(A1-A5) from Appendix A of the same paper). To derive this relation, the dust
geometry of the \cite{Pop11} model was considered.
\begin{eqnarray}\label{eq:Mdust}
\tau_{B}^{f}=K\frac{M_{dust}}{R_{s,d}}
\end{eqnarray}
where $K = 1.0089pc^{2}/kg$ is a constant containing the details of the dust geometry and the spectral emissivity of the \cite{Wei01} model, while $R_{s,d}$ is the scale-length
of the disk, expressed in kpc.

However, one of the differences with respect to Paper I approach is that, for the elliptical galaxies, the dust opacity and dust mass were not derived because the relations used previously
for spiral galaxies were not calibrated for the specific geometry of this type of galaxies. Thus, the correlation 
from Eq.~\ref{eq:Grootes} was discovered by analysing a sample of nearby spiral galaxies, while the relation in Eq.~\ref{eq:Mdust}
relies on the dust geometry of the \cite{Pop11} model (which considers that the diffuse dust in the disc is distributed axisymetrically in two exponential discs), also
calibrated on a small sample of nearby spiral galaxies. Taking into account these considerations, we only used the aforementioned relations (Eqs. (10)-(11)) for the 
spirals and lenticulars in our enlarged sample. 

To correct all the parameters involved in the bulge and ETG scaling relations, we used again the method developed and presented in \cite{Pas13a} and \cite{Pas13b}. More
specifically, we used the whole chain of corrections presented in Eqs.~(4-13) from \cite{Pas13a} and Eqs.~(3-13) from \cite{Pas13b} for the bulges and lenticular galaxies,
together with all the numerical results (given in electronic form as data tables at CDS) to correct the measured parameters for 
projection (inclination), dust and decomposition effects, in order to obtain their dust-free, intrinsic values. For the elliptical galaxies however, we used just the 
projection corrections together with the ones for projection effects in the decomposition process (a part of the third term in the chain of corrections). The reason for
this is that, as we mentioned earlier, the dust opacity was not calculated with \cite{Gro13} relation and thus, dust effects on the structural and photometric parameters of 
the ellipticals could not be estimated.\\
However, as it was shown in \cite{Pas13a} and \cite{Pas13b} (see Figs. 5,6, 19 \& 22 in \cite{Pas13a}), the projection
effects on bulge parameters are far more significant that those for the disc, having a stronger influence than the dust effects over the measured structural parameters of bulges. Therefore, 
these effects should be taking into account for elliptical galaxies, even tough the dust effects are not quantified or considered.\\
Due to the fact that the numerical corrections are a function of wavelength, dust opacity and/or bulge-to-disc ratio, we used the values of $\tau_{B}^{f}$ (where available),
and $B/D$ already derived individually for each galaxy and did all the needed interpolations to obtain the final values for the structural and photometric parameters. 
As a result, $B/D$, $n$, $R_{e,b}$, $\mu_{0,b}$, $\mu_{e,b}$, $<\mu_{e,b}>$, $m_{app,b}$, $M_{abs,b}$ were fully corrected. Thus, we rewrite Eqs.~\ref{eq:SBe_b}-\ref{eq:Mabs_b} 
to determine the intrinsic (corrected) values for these parameters:
\begin{eqnarray}\label{eq:SB0_b_corr}
 \mu_{0,b}^{i}=\mu_{0,b}-A_{ext}+A_{dim}
\end{eqnarray}
where $A_{ext}$ is the attenuation due to the foreground galactic extinction (taken from NED, as in Schlafly \& Finkbeiner 2011
recalibration of the Schlegel et al. 1998 infrared based dust map), and $A_{dim}=-2.5\log(1+z)^3$ is the attenuation due to cosmological redshift dimming, per unit frequency interval,\\
\begin{eqnarray}\label{eq:SBe_b_corr}
 \mu_{e,b}^{i}=\mu_{0_b}^{i}+2.5\times\kappa_{n}/log(10)
\end{eqnarray}
\begin{eqnarray}\label{eq:avg_SBe_b_corr}
 <\mu_{e,b}^{i}>=\mu_{e,b}^{i}-2.5\log([\frac{n^{i}\exp(\kappa_{n})\Gamma(2n^{i})}{\kappa_{n}^{2n^{i}}}]
\end{eqnarray}
\begin{eqnarray}
 m_{b}^{i}=<\mu_{e,b}^{i}>-2.5\log(2\pi(R_{e,b}^{i})^2Q_{b})
\end{eqnarray}
\begin{eqnarray}\label{eq:Mabs_b_corr}
 M_{b}^{i}=m_{b}^{i}-25-5\log(d_{gal}/Mpc)
\end{eqnarray}
All the galaxies in our enlarged sample are at low redshift and therefore we did not apply K-corrections or evolutionary ones. Correspondingly, the correction due to 
cosmological redshift dimming is also quite small, in the range $0.01-0.05$ mag. 
The values for all the previously mentioned parameters are shown in Tables \ref{tab:photo_struct} and \ref{tab:dust}, for the whole sample of spirals and early-type galaxies.
\begin{table*}
\begin{center}
\caption{\label{tab:photo_struct} The photometric and structural parameters for the bulges of late-type (\textbf{LTGs}) and early-type (\textbf{ETGs}) galaxies. The 
columns represent: (1) - galaxy name; (2) - the intrinsic bulge-to-disc ratio; (3) - the observed bulge axis-ratio; (4), (5) - the observed and intrinsic bulge effective
radii; (6-8) - the observed, the intrinsic bulge effective surface brightness, and the standard deviation; (9-11) - the observed, the intrinsic mean effective surface 
brightness, and the standard deviation; (12-14)- the observed, the intrinsic bulge absolute magnitude, and the standard deviation: (15), (16) - observed and intrinsic
bulge S\'{e}rsic index. In square brackets we have the units in which these quantities are expressed.}
\begin{tabular}{{r|r|r|r|r|r|r|r|r|r|r|r|r|r|r|r}}
 \hline \hline
 $Galaxy$  & $(B/D)^{i}$ & $Q_{b}$ & $R_{e,b}$ & $R_{e,b}^{i}$ &  $\mu_{e,b}$ & $\mu_{e,b}^{i}$ & $\sigma_{\mu_{e,b}}$ & $<\mu_{e,b}>$ & $<\mu_{e,b}^{i}>$ & $\sigma_{<\mu_{e,b}>}$ & $M_{b}$ & $M_{b}^{i}$ & $\sigma_{M_{b}}$ & $n$ & $n^{i}$\\\\
   &   &  &  [kpc]  &  [kpc] &  $[\frac{mag.}{arcsec^{2}}]$ & $[\frac{mag.}{arcsec^{2}}]$  &  & $[\frac{mag.}{arcsec^{2}}]$ & $[\frac{mag.}{arcsec^{2}}]$ &  &  [mag.] & [mag.] &  &   \\
   (1)  & (2)  &  (3)  &  (4) &  (5)  &  (6)  &  (7)  &  (8)  &  (9)  &  (10)  &  (11)  & (12)  &  (13)  &  (14)  &  (15)  &  (16)  \\
 \hline 
   \textbf{LTGs}\\
NGC0024  &    0.00  &   0.00  &   0.00  &  0.00  &  0.00 &   0.00  &  0.00  &   0.00  &  0.00  &  0.00  &    0.00  &    0.00  &   0.00  &  0.00  &  0.00\\
NGC0628  &    0.03  &   0.92  &   0.45  &  0.38  & 20.19 &  19.92  &  0.24  &  19.42  & 18.89  &  0.46  &  -17.34  &  -17.49  &   0.58  &  1.17  &  1.45\\
NGC2841  &    0.25  &   0.66  &   0.69  &  0.63  & 19.06 &  18.98  &  0.23  &  18.17  & 17.65  &  0.45  &  -19.13  &  -19.47  &   0.61  &  1.50  &  2.01\\
NGC2976  &    0.00  &   0.00  &   0.00  &  0.00  &  0.00 &   0.00  &  0.00  &   0.00  &  0.00  &  0.00  &    0.00  &    0.00  &   0.00  &  0.00  &  0.00\\
NGC3031  &    0.97  &   0.70  &   1.47  &  0.91  & 20.84 &  20.54  &  0.20  &  19.56  & 18.37  &  0.41  &  -19.45  &  -19.63  &   0.51  &  3.20  &  4.56\\
NGC3190  &    0.31  &   0.27  &   1.53  &  1.26  & 18.54 &  18.40  &  0.27  &  18.11  & 17.83  &  0.52  &  -19.96  &  -19.82  &   0.61  &  0.56  &  0.66\\
NGC3621  &    0.03  &   0.53  &   1.15  &  0.40  & 21.29 &  20.98  &  0.31  &  21.19  & 20.88  &  0.62  &  -17.01  &  -15.04  &   0.70  &  0.15  &  0.21\\
NGC3938  &    0.03  &   0.94  &   0.45  &  0.35  & 20.22 &  20.11  &  0.29  &  19.60  & 19.23  &  0.52  &  -17.19  &  -16.99  &   0.82  &  0.86  &  1.08\\
NGC4254  &    0.21  &   0.79  &   1.65  &  1.54  & 21.97 &  21.80  &  0.21  &  20.93  & 20.32  &  0.42  &  -18.46  &  -18.93  &   0.75  &  2.02  &  2.64\\
NGC4450  &    1.01  &   0.62  &   2.18  &  3.86  & 21.75 &  21.62  &  0.21  &  20.40  & 19.17  &  0.41  &  -19.33  &  -21.82  &   0.64  &  3.63  &  5.31\\
NGC4594  &    4.81  &   0.76  &  21.31  &  6.48  & 25.87 &  25.66  &  0.20  &  24.38  & 22.62  &  0.40  &  -20.53  &  -19.71  &   0.44  &  4.76  &  7.18\\
NGC4736  &    1.08  &   0.90  &   0.27  &  0.20  & 17.89 &  17.82  &  0.23  &  16.97  & 16.47  &  0.44  &  -18.66  &  -18.50  &   0.59  &  1.60  &  2.11\\
NGC4826  &    0.04  &   0.66  &   0.12  &  0.07  & 18.28 &  18.12  &  0.32  &  17.70  & 17.27  &  0.55  &  -15.82  &  -15.13  &   0.83  &  0.78  &  0.99\\
NGC5033  &    0.35  &   0.40  &   1.76  &  0.74  & 20.18 &  20.10  &  0.23  &  19.34  & 18.85  &  0.44  &  -19.46  &  -18.06  &   0.69  &  1.34  &  1.78\\
NGC5055  &    0.15  &   0.63  &   1.32  &  0.59  & 21.21 &  21.13  &  0.23  &  20.47  & 20.08  &  0.45  &  -18.19  &  -16.81  &   0.78  &  1.10  &  1.40\\
NGC5474  &    0.19  &   0.77  &   1.04  &  1.08  & 23.82 &  23.77  &  0.22  &  22.86  & 22.42  &  0.43  &  -15.51  &  -16.03  &   0.62  &  1.72  &  2.24\\
NGC7331  &    0.25  &   0.40  &   1.25  &  1.07  & 19.35 &  18.99  &  0.25  &  18.81  & 18.26  &  0.49  &  -19.24  &  -19.44  &   0.62  &  0.71  &  0.86\\
NGC7793  &    0.01  &   0.85  &   0.03  &  0.02  & 17.73 &  17.66  &  0.60  &  16.90  & 16.58  &  0.81  &  -13.76  &  -12.87  &   1.44  &  1.33  &  1.62\\
\hline
  \textbf{ETGs}\\ 
NGC0584  &    1.86  &   0.66  &   1.40  &  1.15  & 19.37 &  19.19   & 0.22 &  18.23  & 18.04  &  0.42  &  -20.63  &  -20.38  &   0.62  &  2.50  &  2.49\\
NGC0855  &    0.38  &   0.51  &   0.36  &  0.30  & 21.11 &  20.84   & 0.25 &  20.30  & 19.92  &  0.47  &  -15.31  &  -15.31  &   0.72  &  1.27  &  1.41\\
NGC1404  &    5.21  &   0.75  &   4.87  &  3.99  & 21.39 &  21.31   & 0.20 &  20.00  & 19.94  &  0.40  &  -21.68  &  -21.32  &   0.54  &  3.92  &  3.87\\
NGC3265  &    0.73  &   0.68  &   0.83  &  0.70  & 20.61 &  20.49   & 0.38 &  19.81  & 19.57  &  0.60  &  -17.94  &  -17.80  &   0.86  &  1.26  &  1.41\\
NGC4125  &    3.20  &   0.50  &   4.74  &  3.90  & 21.22 &  21.11   & 0.21 &  19.98  & 19.88  &  0.41  &  -21.22  &  -20.90  &   0.68  &  2.95  &  2.93\\
NGC4552  &    0.69  &   0.90  &   1.79  &  1.47  & 20.87 &  20.70   & 0.21 &  19.59  & 19.42  &  0.41  &  -20.13  &  -19.87  &   0.59  &  3.25  &  3.21\\
NGC1377  &    1.62  &   0.49  &   1.95  &  1.62  & 22.22 &  22.08   & 0.23 &  21.35  & 21.13  &  0.44  &  -17.90  &  -17.71  &   0.94  &  1.44  &  1.55\\
NGC1482  &    1.08  &   0.55  &   3.67  &  3.06  & 21.88 &  21.71   & 0.21 &  20.92  & 20.69  &  0.42  &  -19.82  &  -19.65  &   1.25  &  1.74  &  1.83\\
NGC1705  &    1.29  &   0.85  &   0.12  &  0.10  & 18.60 &  18.56   & 0.27 &  17.28  & 17.26  &  0.48  &  -16.53  &  -16.04  &   0.98  &  3.46  &  3.37\\
NGC3773  &    0.62  &   0.57  &   0.22  &  0.19  & 17.97 &  17.84   & 0.41 &  17.47  & 17.02  &  0.65  &  -17.18  &  -17.28  &   1.78  &  0.66  &  0.88\\
NGC5866  &    0.46  &   0.67  &   8.44  &  7.55  & 24.13 &  24.05   & 0.26 &  24.06  & 23.32  &  0.53  &  -18.71  &  -19.21  &   0.68  &  0.27  &  0.57\\
\hline
 \end{tabular}
 \end{center}
\end{table*}

\begin{table*}
\caption{\label{tab:dust} Dust masses and dust opacities, derived using Eqs.~(1) and (5) from \protect\cite{Gro13}, for the early-type galaxies in our sample. 
The different columns represent: (1) - galaxy name; (2) - B band face-on dust optical depth; (3) - stellar mass surface
densities; (4) - corrected stellar mass surface densities; (5) - stellar masses taken from: $a$ - \protect\cite{Remy15}, $b$ - \protect\cite{Ski11},
$c$ - \protect\cite{Wilson13}, $d$ - \protect\cite{Zib11}, $e$ - \protect\cite{Grossi15};
(6) - dust masses; (7) - corrected dust masses; (8)-(11) - standard deviation for $\protect \tau_{B}^{f}$, $\protect \mu_{*}$, $M_{*}$ and $\protect M_{dust}$. The errors
for the corrected quantities are the same and thus not given here. In square brackets we have the units in which these quantities are expressed. All quantities except
dust optical depth are given in decimal logarithm unit scale.}
 \begin{tabular}{{r|r|r|r|r|r|r|r|r|r|r}}
  \hline \hline
 $Galaxy$ &  $\tau_{B}^{f}$ &  $log(\mu_{*})$ & $log(\mu_{*}^{i})$ & $log(M_{*})$ & $log(M_{dust})$ & $log(M_{dust}^{i})$ & $\sigma_{\tau_{B}^{f}}$  & $\sigma_{\mu_{*}}$ & $\sigma_{M_{*}}$ & $\sigma_{M_{dust}}$\\
          &                 &  $[M_{\odot}/kpc^{2}]$ & $[M_{\odot}/kpc^{2}]$  &  $[M_{\odot}]$ &  $[M_{\odot}]$  &  $[M_{\odot}]$  &   &  $[M_{\odot}/kpc^{2}]$  & $[M_{\odot}]$  & $[M_{\odot}]$  \\
  \hline
NGC0584 &  --  &  8.50  &  9.02 &  $10.87^{a}$  &  --  &  --  &  --  &  0.20  &  0.20  &  --\\   
NGC0855 &  --  &  7.75  &  8.19 &  $ 9.20^{a}$  &  --  &  --  &  --  &  0.21  &  0.21  &  --\\    
NGC1404 &  --  & 10.21  & 10.80 &  $10.85^{b}$  &  --  &  --  &  --  &  0.14  &  0.13  &  --\\    
NGC3265 &  --  &  6.36  &  6.80 &  $ 9.55^{a}$  &  --  &  --  &  --  &  0.21  &  0.21  &  --\\    
NGC4125 &  --  &  8.73  &  9.27 &  $11.38^{c}$  &  --  &  --  &  --  &  0.11  &  0.11  &  --\\    
NGC4552 &  --  &  7.54  &  8.10 &  $10.97^{d}$  &  --  &  --  &  --  &  0.10  &  0.10  &  --\\   
NGC1377 &  0.01  &  5.98  &  6.43 &  $ 9.47^{a}$  &  6.34  &  6.34  &  0.01  &  0.21  &  0.21  &  0.24\\   
NGC1482 &  0.08  &  6.69  &  7.16 &  $ 9.99^{b}$  &  6.94  &  6.92  &  0.02  &  0.11  &  0.11  &  0.12\\    
NGC1705 &  0.90  &  7.64  &  8.22 &  $ 8.19^{e}$  &  5.26  &  5.13  &  0.16  &  0.07  &  0.04  &  0.12\\    
NGC3773 &  0.21  &  7.08  &  7.48 &  $ 8.31^{b}$  &  5.31  &  5.36  &  0.09  &  0.16  &  0.16  &  0.19\\   
NGC5866 &  5.07  &  8.31  &  8.70 &  $10.02^{b}$  &  7.17  &  7.22  &  1.19  &  0.09  &  0.09  &  0.11\\    
\hline
 \end{tabular}
\end{table*}

\subsection{Estimation of errors}\label{sec:errors}
 
As previously mentioned in \cite{Pas20}, the best-fit parameters given by GALFIT suffer from an underestimation of uncertainties, which was shown in \cite{Hau07}. To 
estimate the systematic errors on the main photometric parameters involved in this study, namely the ones that characterise the bulges of spiral galaxies and the ETG 
galaxies, we ran a new set of fits, for a few sampled galaxies (both late-type and early-type). In this process, we fixed the sky value to the one found initially by 
GALFIT and added $\pm1\sigma$, or $\pm3\sigma$ ($\sigma$ being the uncertainty in the sky level), leaving free the parameter of interest (either bulge effective radius, 
axis-ratio or S\'{e}rsic index) while all other parameters were also fixed to the values found by GALFIT. This approach of error estimation of bulge parameters 
was also used previously by \cite{Gao20} and \cite{Gao19}, following from \cite{Gao17} and \cite{Hua13a}.
The systematic errors in the bulge effective radii were within the range 1-10 pixels (1-3 arcsecs). For the bulge axis-ratios, the uncertainties were less significant, of
up to $0.01$. For the S\'{e}rsic index, the uncertainties observed were more significant as this parameter is quite sensitive to the choice of the sky or innacurare star
masking, and in general is determined with less precision compared with other photometric parameters. Thus, the uncertainties were in the range $0.1-0.75$ which corresponds
to 0.04-0.2 dex for log$(n_{b})$. The error over $d_{gal}$ (measured distance to the galaxy) was taken from NED. Having the flux uncertainties already shown in Table~\ref{tab:photo_fluxes},
we performed propagation of errors in Eqs.~\ref{eq:SBe_b}-\ref{eq:Mabs_b_corr} to calculate the standard deviations ($\sigma$) for all the required parameters. The uncertainties
are shown in Table~\ref{tab:photo_struct}.
\begin{figure*}
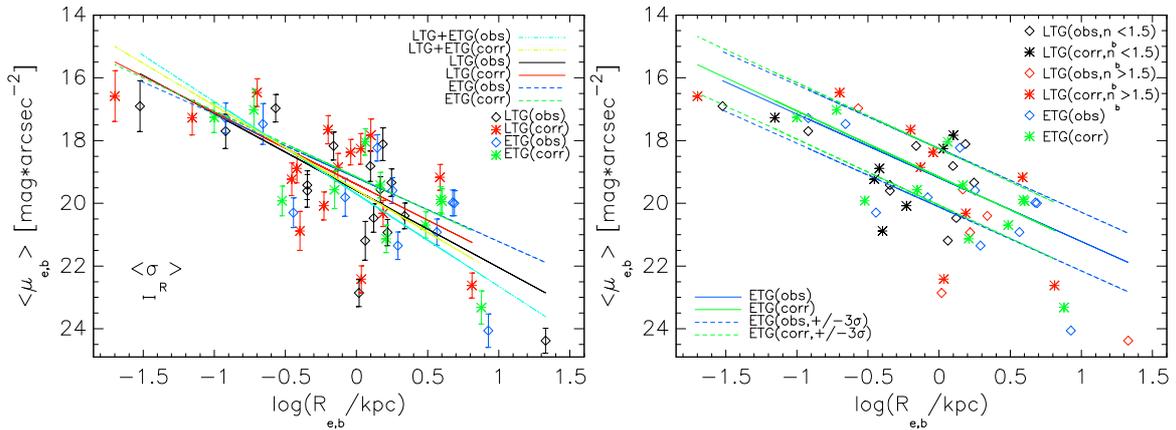

 \begin{center}
  \includegraphics[scale=0.45]{Fig_1a.epsi}
  \hspace{-0.1cm}
  \includegraphics[scale=0.450]{Fig_1b.epsi}
  \caption{\label{fig:Kormendy_relation} \textit{Left panel} The Kormendy relation for the ETGs and LTGs in our sample. The observed average bulge surface brightness 
  ($<\mu_{e,b}>$) values for the bulges of spirals are represented with black diamonds, while the corrected ones are plotted with red stars. The solid black and red lines are
  the best-fit to the data points obtained through a linear regression. Similarly, for the early-type galaxies, the blue diamonds show the observed (measured) $<\mu_{e,b}>$ 
  values, while green stars correspond to the corrected (intrinsic) ones. Similar to the spirals, green and blue dashed lines are the best-fit Kormendy relations for the ETG 
  galaxies. Additionally, we plot with yellow and cyan dash-dotted lines the best-fit to all the data points. The error bars represent the standard deviations. The average 
  uncertainty of $R_{e,b}$ is shown in the lower left side. \textit{Right panel} The same relation, but with the sample of spiral galaxies now divided according
  with the bulge S\'{e}rsic index - black diamonds and stars show the observed and intrinsic values for spiral galaxies with $n_{b}<1.5$, while red diamonds and stars 
  correspond to the ones with $n_{b}\geq1.5$. We overplot the best-fits relation for early-type galaxies with blue and green solid lines (observed and intrinsic) and
  the corresponding $\pm3\sigma$ boundaries with dashed blue and green lines. For comparing purposes, we additionally plot the data points for ETGs, just as in the
  left panel (same symbols, not divided by $n_{b}$). The legends overplotted in both panels illustrate the correspondence of each symbol and line to the data.} 
\end{center}
\end{figure*} 
 
\section{Results}\label{sec:results}

In this section, the main results of this study are presented, the scaling relations obtained for the bulges of the spiral galaxies and the early type galaxies in our sample.\\

\subsection{The Kormendy relation}\label{sec:Kormendy}

We present first, the relation between the mean bulge surface brightness ($<\mu_{e,b}>$) and bulge effective radius, the photometric projection of the Fundamental Plane of
elliptical galaxies (Djorkovsky \& Davis 1987, Dressler et al. 1987), also known as the Kormendy relation (Kormendy 1977), in the left-hand panel of Fig.~\ref{fig:Kormendy_relation}.
We derive the general relation, for the whole sample of late-type and early type galaxies (dash-dotted cyan and yellow lines for the measured and corrected relation), but also separately for
bulges of the spiral galaxies and the early-type ones (ellipticals+lenticulars). We do this as a means to verify if there are significant differences between the characteristics
of the relation for bulges of spirals and the ones corresponding to early-type galaxies and if these two morphologically distinct categories occupy a different locus on the
$<\mu_{e,b}>-R_{e,b}$ relation. Moreover, both the observed (black / blue diamonds for LTG/ETG) and intrinsic (red / green stars for LTG/ETG) Kormendy relation are shown.\\
The best-fit for all cases is obtained through a linear regression procedure, having the slope and the intercept as coefficients of the relation $<\mu_{e,b}>=a+b\times\log(R_{e,b})$, 
and given in Table~\ref{tab:Kormendy_coeff}. \\
Looking at the observed best-fit Kormendy relation for bulges of spiral galaxies compared with corresponding one for the early-type galaxies, one can notice differences in
the slope of the relation, with bulges residing on a slightly steeper slope relation than ETGs. This could be explained in part by the changes induced by dust and projection
effects, which have a larger extent for bulges in spirals. We remind the reader that dust effects were not quantified for the ellipticals in our sample. This can also be
checked if we compare the pair of observed to intrinsic relations for bulges (solid black and red lines) and early-type galaxies (dashed blue and green lines). Thus, we notice
a more significant change in the slope and intercept for the former ones (see also Table~\ref{tab:Kormendy_coeff}). The steepest relation is the general one, for the whole
sample, with an observed slope of $2.94\pm0.21$, consistent with other recent studies (e.g. Sachdeva et al. 2020, Gao et al. 2020), while the intrinsic relation suffering a more 
important decrease in slope due to the aforementioned biases compared with the separated relations.\\ 
\cite{Gad09} and \cite{Gao20} performed a multi-component structural analysis on a large sample of elliptical and disc galaxies, using the Kormendy relation to separate
classical and pseudo-bulges. Thus, they considered the best-fit Kormendy relation of ellipticals and classified pseudo-bulges as those bulges that are situated 
below the $3\sigma$ boundary of the aforementioned relation. They showed that this is more precise means
for bulge classification than S\'{e}rsic indices. Moreover, in Fig. 8 of \cite{Gad09} and Figs. 5\&6 of \cite{Gao20}, the authors showed that when the bulges are divided
on the basis of their S\'{e}rsic indices in the Kormendy relation, there is a pronounced overlap of the distribution of bulges and pseudo-bulges. We have not explicitely separated our
sample into galaxies with classical or pseudo-bulges. Nevertheless, we tested this result by dividing the spiral galaxies sample based on our derived 
S\'{e}rsic indices, into galaxies with $n_{b}<1.5$ bulges, and the rest with $n_{b}\geq1.5$ (as we would try to classify the sample into galaxies containing a classical or
a pseudo-bulge), to check how the bulges classified in this way would be situated with respect to the Kormendy relation of the ETGs.
We show this in the right panel of Fig.~\ref{fig:Kormendy_relation}, where the black diamonds correspond to the observed values for spiral galaxies with low S\'{e}rsic index 
$n_{b}>1.5$, while the red diamonds show the other category of spiral galaxies (with $n_{b}>1.5$). In a similar way, we show the intrinsic values, with black and red stars, 
respectively. We overplot the best-fit observed and intrinsic Kormendy relation for ETGs from the left-hand panel in solid blue and green lines, while the corresponding
dashed lines represent the $\pm3\sigma$ boundaries over the zero-point (the slope remaining fixed). Additionally, for comparing purposes, we overplot the data points
corresponding to the sample of early-type galaxies, just as in the left panel - not divided by a $n_{b}$ value (same symbols and color scheme). One can see that most of the 
high S\'{e}rsic index LTGs (with classical bulges) are situated in the region where most of the early-type galaxies reside. Nevertheless, comparing the two plots, using
the available morphological classification and taking into account that our sample is rather small (but representative), we cannot clearly observe a similar separation as
in Fig.~1 of \cite{Gao20}, where most of their classified pseudo-bulges lie beyond the 3$\sigma$ boundary of their ellipticals Kormendy relation. However, we do 
notice, as in Fig.~8 in \cite{Gad09} and Fig.~5 in \cite{Gao20}, that high- and low-S\'{e}rsic index bulges significantly overlap in the $<\mu_{e,b}>-R_{e,b}$ relation.
This is also the case when the sample of LTGs is divided using a $n_{b}=2.0$ threshold. This result is in contrast with that found by \cite{Neu17}, who
analysed a small sample of unbarred disc galaxies and found that using the Kormendy relation of ETGs and separating their bulges using a $n_{b}=1.5$ threshold produces a clear
separation of ellipticals, low- and high-S\'{e}rsic index bulges, with almost no overlap. The authors also found pseudo-bulges to be located below the 2$\sigma$ boundary of the relation.
In contrast with \cite{Gad09} and \cite{Gao20}, \cite{Cost17} found the Kormendy relation not reliable for separating bulge populations due to the large scatter and the overall
dependence of the best-fit coefficients of this relation on the absolute magnitude range considered. Moreover, \cite{Fish16} found that selecting bulges solely by their 
position in the $<\mu_{e,b}>-R_{e,b}$ space does not clearly separate bulges, while adding a $n_{b}=2.0$ selection threshold to the Kormendy relation produces a cleaner separation
of classical and pseudo-bulges.\\
Thus, according with our findings, using the Kormendy relation of early-type galaxies in combination with the S\'{e}rsic index of bulges does not provide a relevant
and conclusive classification of bulge morphology. This happens because the S\'{e}rsic index is not a good bulge morphology discriminator, and using it in combination
with an apparently more accurate criterium (the Kormendy relation for ETGs) does not produce conclusive results. Thus, we want to emphasize that a threshold in the 
bulge / galaxy S\'{e}rsic alone is not sufficient for an accurate separation between classical and pseudo-bulges, as it may lead to errors and misclassifications. Using $B/T$ 
additionally would give us a more clear picture, as one can see that bulges with higher S\'{e}rsic index are situated in general at higher effective radii and $<\mu_{e,b}>$ 
on the Kormendy relation (right panel of Fig.~\ref{fig:Kormendy_relation}).\\ 
In Table~\ref{tab:Kormendy_coeff} we also show the derived scatter (rms in the $<\mu_{e,b}$ direction) for all the plotted relations and the corresponding correlation coefficients. 
The scatter for the general relation and the one for the spiral galaxies is comparable with small differences between the values for the measured and intrinsic relations. In
the case of early-type galaxies, the scatter is larger, mainly due to the reduced number of galaxies analysed, compared with the overall sample and the number of spirals. As
expected for a well-established relation, the Pearson coefficients are close to 1.0, indicating a near perfect correlation. The coefficents are lower for the ETGs, which may 
also be explained by the above argument for the rms.\\
\begin{figure}
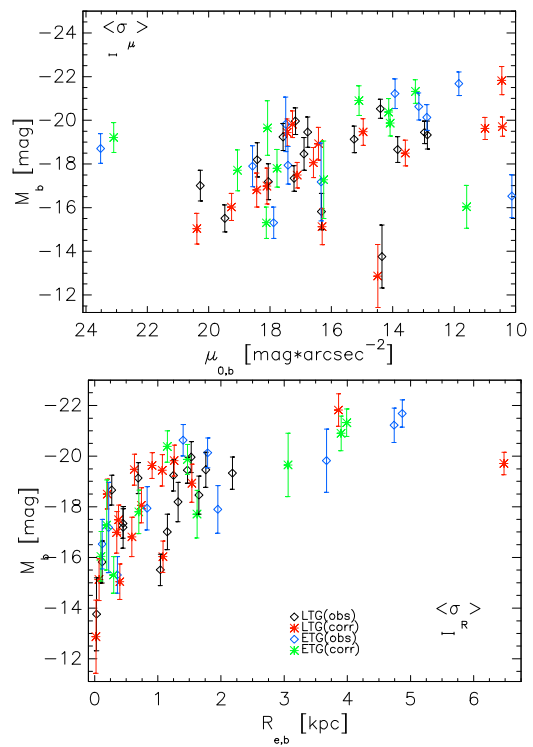

\begin{center}
 \includegraphics[scale=0.39]{Fig_2a.epsi}
 \includegraphics[scale=0.39]{Fig_2b.epsi}
 \caption{\label{fig:Mb_vs_Re_mu0} \textit{Upper panel} Bulge absolute magnitudes as a function of bulge central surface brightness, $\mu_{0,b}$. The average 
  uncertainty of $\mu_{0,b}$ is shown in the upper left corner. \textit{Lower panel} Bulge absolute magnitudes as 
 a function of bulge effective radii. The average uncertainty of $R_{e,b}$ is shown in the lower right corner. The symbols and color scheme (common for both panels and 
 illustrated in the legend overplotted in the bottom panel) are as follows: 
 black and blue diamonds represent the observed values for bulges of spiral galaxies and the early-type ones;
 red and green stars correspond to the intrinsic values. The error bars represent the standard deviations.}
 \end{center}
\end{figure}
\subsection{Other bulge scaling relations}\label{sec:others}

The bulge absolute magnitudes are plotted against the bulge central surface brightness in the upper panel of Fig.~\ref{fig:Mb_vs_Re_mu0}. The linear continuous trend is noticed 
for both the observed and corrected data points, without any real, noticeable separation for spirals from early-type galaxies, as also shown in \cite{Gra2013} and \cite{Guz03}. This relation
is in close connection with another linear relation, namely the $M_{b}-n_{b}$ one, which will be shown in the next section, dedicated to the black-hole scaling relations. The apparent
small gaps in the data at $\mu_{0,b}=16.0$ and $13.0$ are mostly due to our sample size. In the lower panel of the same figure, the bulge absolute magnitude is plotted as a function
of the effective radius, with the symbols and colors being the same as in the previous plot. Although we do not attempt to fit the data, we can see that the relation $M_{b}-R_{e,b}$
is curved, unlike the previous one, as are all relations which involve effective quantities (e.g. $R_{e}, \mu_{e}$ or $<\mu_{e}>$), a fact pointed out by \cite{Gra2013}.
\begin{figure}
\begin{center}
 \includegraphics[scale=0.39]{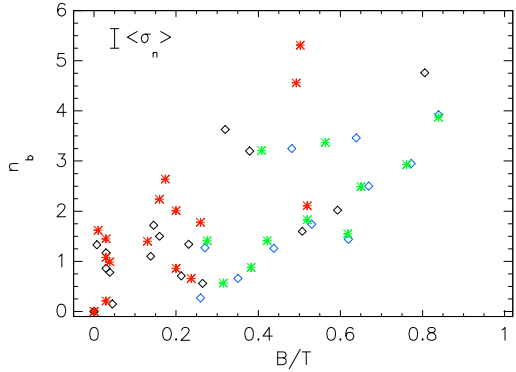}
 \caption{\label{fig:nb_vs_BT} The S\'{e}rsic index of the bulges of spirals and early-type galaxies vs. $B/T$ (bulge-to-total flux ratio). The average 
  uncertainty of $n_{b}$ is shown in the upper left corner, while the average uncertainty of $B/T$ is $0.025$.
  The symbols and color scheme are the same as in Fig.~\ref{fig:Mb_vs_Re_mu0}.}
\end{center}
\end{figure}

\begin{table}
\begin{center}
 \caption{\label{tab:Kormendy_coeff} The linear regression best-fit parameters for the Kormendy relation, for the late type and early type galaxies shown in the left panel of Fig.~\ref{fig:Kormendy_relation},for both the
 measured (observed) and corrected (intrinsic) ones: $a$ - the intercept, $b$ - the slope. We also present here in the last two columns, the scatter of the relations, $\sigma$, and Pearson correlation
 coefficient, $r^{corr}$. }
 \begin{tabular}{{r|r|r|r|r}}
  \hline \hline
  Kormendy relation      	&  $a$  &  $b$  &  $\sigma$ (rms)  &   $r^{corr}$\\
 \hline
 obs., LTG+ETG    &  $19.70\pm0.12$  &  $2.94\pm0.21$ &  0.59  &  0.90   \\
 intrin., LTG+ETG &  $19.57\pm0.13$  & $2.69\pm0.21$ &   0.63  &  0.87  \\
 obs., LTG    &  $19.59\pm0.18$  & $2.45\pm0.33$  &  0.58  &  0.98  \\
 intrin., LTG   & $19.39\pm0.20$ & $2.29\pm0.32$  &  0.56  &  0.91  \\
 obs., ETG   &  $19.19\pm0.31$  & $2.03\pm0.56$  & 0.80  &  0.83  \\
 intrin., ETG   &  $19.16\pm0.30$ & $2.11\pm0.56$ & 0.79  & 0.84   \\
  \hline               
 \end{tabular}
 \end{center}
\end{table}
In figure \ref{fig:nb_vs_BT}, the bulge S\'{e}rsic index is plotted against the bulge-to-total flux ratio ($B/T$). The black and blue diamomds represent the observed values
of the spiral bulges and those of the early type galaxies, while red and green stars show the corresponding intrinsic values. One can see a quite clear separation in the 
$n_{b}-B/T$ plane for bulges of spiral galaxies and those of ETG, irrespective of a few outliers. Thus, the early-type ones occupy the area of high $B/T$ and moderate to 
high values of the S\'{e}rsic index, as expected. This holds true for both the measured and intrinisc parameters. A similar result was observed by \cite{Gad09} by doing a
structural analysis of a much larger sample of galaxies from the SDSS survey. Both \cite{Gad09} and \cite{Fish08} have also shown that a correlation exists between 
$n_{b}$ and $B/T$ for classical bulges, while this is not present in the case of pseudo-bulges. Although our sample is smaller, we can confirm this correlation, even though 
we did not divide our sample into pseudo- and classical bulges. \cite{Lau10} observed a similar separation in the $n_{b}-B/T$ plot between the regions ocupied by 
lenticulars and Sa-Sc galaxies, and determined that a correlation exists between these two measured quantities. The fact that late-type and early-type galaxies reside in                     
distinct areas in the $n_{b}-B/T$ plane confirms this relation as a useful and accurate criterion to morphologically classify galaxies, as long as $n_{b}$ and $B/T$ are
used in combination, and with the inherent biases taken out from measurements. On a different note, \cite{Cost17} and \cite{Cost18} found that bulges with
low S\'{e}rsic index found in galaxies with low $B/T$ are not naturally disc-like, rotationally supported components (as pseudo-bulges), and cand have a variety of intrinsic
shapes, unlike classical, spheroid-like bulges.\\
Here we also note the average bulge S\'{e}rsic index (measured and intrinsic) for the spirals - $<n_{b}>=1.47$  \&  $<n_{b}^{i}>=2.01$, while for ETGs the respective
values are $<n_{b}>=2.06$  \&  $<n_{b}^{i}>=2.14$. These values are consistent with those found in other studies such as those of \cite{Lau07} or \cite{Gra01}.\\
\begin{figure}
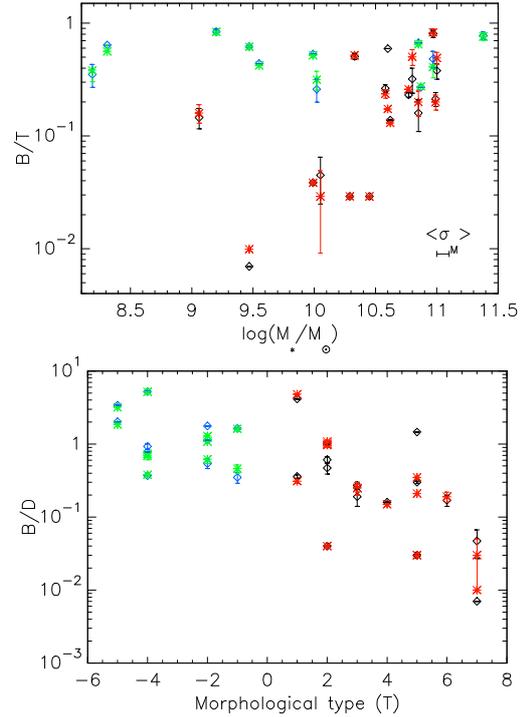

\begin{center}
 \includegraphics[scale=0.38]{Fig_4a.epsi} 
 \hspace{-0.1cm}
 \includegraphics[scale=0.38]{Fig_4b.epsi} 
 \caption{\label{fig:BD_type_rel} \textit{Upper panel} Bulge-to-total ($B/T$) ratio of the sample, as a function of stellar mass. The average 
  uncertainty of $M_{\star}$ is shown in the lower right corner. \textit{Lower panel} Bulge-to-disc ratio 
 versus morphological type, $T$. The symbols and color legend are the same as in Fig.~\ref{fig:Mb_vs_Re_mu0}. 
 The error bars represent the standard deviations.}
 \end{center}
\end{figure}
Another relation that can be used when trying to separate galaxies with different morphologies is the one displayed in the upper panel of Fig.~\ref{fig:BD_type_rel}, namely
the bulge-to-total flux ratio versus the galaxy stellar mass, $M_{\star}$. The $B/T$ values have been corrected for inclination and dust effects for the late-type and 
lenticulars, while for those corresponding to ETGs, only the inclination (projection) corrections were applied. The black and blue triangles are the observed values for 
the late-type and early type galaxies, while the red and green stars represent the corrected values. It can be seen the clear difference in the trends followed by the two
classes of galaxies. The late type bulge-to-total ratios show an increase with stellar mass from very low, almost bulgeless galaxies, to values similar to early-type galaxies.
In contrast, the early-type  (ellipticals and lenticulars) show less of a trend, being almost constant with stellar mass, with values fluctuating around $B/T=0.5$. The corrections 
introduced by the aforementioned effects do not produce important changes in the trend seen for the observed versus the one corresponding to the intrinsic values. Overall,
we consider $B/T - M_{\star}$ relevant when trying to divide galaxies into late-type and early-type galaxies.\\
In the lower panel of Fig.~\ref{fig:BD_type_rel} we show $B/D$ as a function of the morphological type ($T$) for the spiral galaxies. This relation, sometimes in the 
form $B/T-T$, has been shown in similar studies by \cite{Dong16}, \cite{Lau07}, \cite{Gra08}, \cite{Wein09}, \cite{Lau10}, \cite{Men17} and \cite{Gao19}. As expected, one
can notice an important decrease in the bulge-to-disc ratio when going from early to later morphologies, as the bulge influence on the surface brightness profile of the 
galaxy starts to gradually decrease. The trend seen here is similar with that shown in the previous works mentioned above. One can also notice a certain degree of
scatter in the values of $B/T$. A large scatter has been reported before in studies on larger samples than ours, with better statistics, such as the ones of \cite{Wein09},
\cite{Lau10} and \cite{Gao19}. This is the reason other authors concluded that due to this systematic scatter in $B/T$ at any Hubble type, one should not use Hubble type to
quantitavely infer $B/D$ values and/or bulge proeminence. Due to the low statistics for each morphological type in our sample, we cannot examine this issue in more detail or 
make a fair comparison of the average value found for bulge-to-total flux ratios or $n_{b}$, for different morpholgies, like S0 or ellipticals. We just note there the respective
values for the average intrinsic $B/T=0.48$ for lenticular galaxies and 0.56 for the ellipticals.\\
As in the upper plot, we can see that dust and inclination effects do not produce great changes in the $B/D$ values for the spiral galaxies, unlike for the rest of the disc
and bulge structural and photmetric parameters.

\subsection{Black-hole scaling relations}\label{sec:black-hole}

In this section, we present an important set of relations, the black-hole scaling relations, essential for understanding the co-evolution of galaxies (more specific their 
bulge component) and their central black holes, which reside at their centre. These relations are likewise important as they can infer predictions on the black
hole masses of galaxies using their bulge structural or photometric parameters. Since dust and inclination can bias some of the parameters involved in these 
relations, the bulge photometric and structural parameters, it's important to see if there are any important changes introduced by these effects in the main characteristics
of these relations (e.g. the slope, the scatter). We have not derived here the black-hole masses. These were compiled from the relevant papers in the literature, and are given
in Table~\ref{tab:black_holes}.

\begin{figure*}
 \begin{center}
  \includegraphics{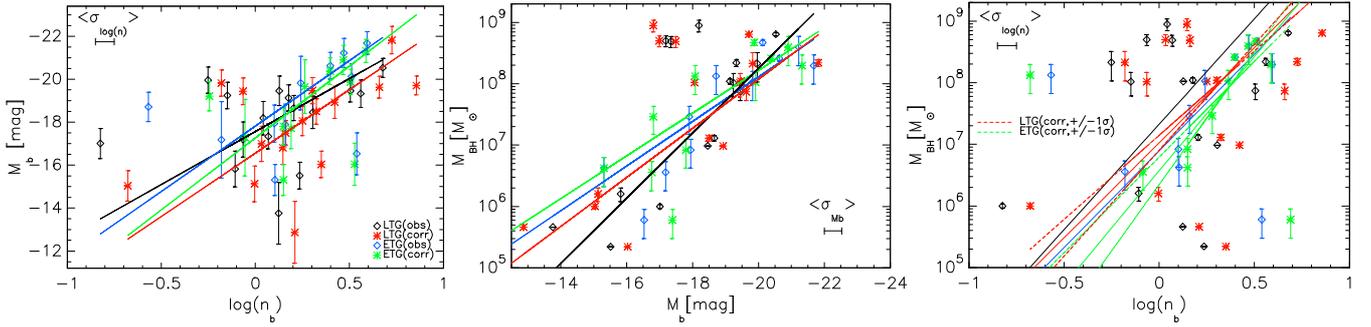}
  \caption{\label{fig:BH_scal_rel} Black-hole scaling relations for the ETGs and LTGs in our sample. The panels are as follows: \textit{Left} Bulge absolute magnitude 
  (luminosity) vs. bulge S\'{e}rsic index; \textit{Middle} Black-hole mass as a function of bulge absolute magnitude; \textit{Right} Black-hole mass plotted against the 
  bulge S\'{e}rsic index. The solid lines represent the best-fit relation to the data points from a linear regression procedure. The symbols and color legend, 
  common for all panels, is shown on the left hand panel. Additionally, in the right-hand panel, the $\pm1\sigma$ limits just for the corrected $M_{BH}-n_{b}$ 
  relations of the LTG and ETG galaxies are overplotted in red and green dashed lines, with the corresponding color and line legend displayed. The error bars represent
  the standard deviations. The average uncertainties of $log(n_{b})$ and $M_{b}$ are overplotted in the upper left and lower right corners.}
\end{center}
\end{figure*}

The relation between the mass of the central black hole ($M_{BH}$) and the stellar light concentration of the galaxy's bulge/spheroidal component, given by 
the S\'{e}rsic index, was first observed by \cite{Gra01}. This was thought to exist as there were already proven correlations which involved the stellar luminosity of the 
bulge / spheroid $L_{b}$ (or alternatively, the bulge absolute magnitude, $M_{b}$), such as the $M_{BH}-L_{b}$ correlation (e.g. Kormendy \& Richstone 1995, Magorian et al. 1998, 
Marconi \& Hunt 2003), or the $L_{b}-n_{b}$ one (e.g. Young \& Currie 1994, Jerjen et al. 2000, Graham \& Guzm\'{a}n 2003). Thus, \cite{Gra07} showed for the first time the
$M_{BH}-n_{b}$ correlation by doing a structural analysis of a small sample of spiral and elliptical galaxies.\\
We present the $M_{b}-\log(n_{b})$ relation in the left-hand panel of Fig.~\ref{fig:BH_scal_rel}, with the symbols and the colors the same as in Fig.~\ref{fig:Kormendy_relation}. 
This is a linear relation, as is the $M_{b}-\mu_{0,b}$ one, presented in the previous section, irrespective of a few outliers. The
trend of the bulge absolute magnitude (or alternatively bulge luminosity, $L_{b}$) is to linearly increase with the concentration of the bulge / spheroid stellar light, for 
both late-type galaxy bulges and early type ones. We do a linear regression procedure to derive the specific parameters of the relation (slope and zero-point). The best-fit 
measured and corrected relations are plotted with black and red solid lines for the spiral galaxies, while the corresponding ones for the ETGs are plotted with blue and green lines. 
Comparing the observed or the corrected relations between bulges of spirals and early-type galaxies, one immediately notices a difference in both the slope and the zero-point of the 
relations, with the ETGs showing a steeper slope relation: $2.44\pm0.39$ vs. $1.97\pm0.16$ for the observed relation, and $2.68\pm0.45$ vs. $2.36\pm0.18$ for the correted one (see Table~\ref{tab:BH_scal_rel_parameters}). These values were obtained
from the initial best-fit parameters of the $M_{b}-n_{b}$ relation, knowing that $M_{b}=-2.5\times\log(L_{b})$, and are thus the best-fit parameters for the $L_{b}-n_{b}$ relation.\\
Different slopes for the ETGs and LTGs were also reported previously in \cite{Gra01}, \cite{Savo13} and \cite{Savo16b}. Our results for the slope of the corrected relation show a
shallower slope than the one found by \cite{Savo16b} - $3.60\pm0.19$, for early-type galaxies (ellipticals+lenticulars), while for the late-types it is steeper than the $1.44\pm0.52$
value derived by the same authors. One can see that dust and inclination effects produce a significant change in the slope and intercept of this relation, especially for the bulges
of late-type galaxies. However, as \cite{Savo16b} performed their multicomponent structural analysis on $3.6\mu$m band images, which are not much affected by dust extinction, dust and inclination
effects alone cannot account for this difference when compared with our intrinsic values. In Section~\ref{sec:discussion}, we are coming back to this issue.\\
In the last two columns of Table~\ref{tab:BH_scal_rel_parameters} we also show the scatter of this relation ($\sigma$) in the $M_{b}$ direction, and the Pearson correlation
coefficients. The minimum number of outliers, situated far-off from the average value, were given minimum weight in the linear regression and thus were practically excluded from these calculations.
However, all the data points were considered in the Pearson correlation coefficients calculation, even though it is known these coefficents are quite sensitive to far-off
values which are prone to high uncertainties, and including them would lead to misinterpretations of the overall trend observed and of the results obtained. 
Comparing the respective values, there are small differences in the scatter, with the intrinsic relation for the 
early-type galaxies having a slightly less scatter than the bulges of spiral galaxies. The values are comparable, though smaller, with those found by \cite{Savo16b}. Similar 
behaviour can be observed for the correlation coefficients, with very small differences between the coefficients for the observed vs. intrinsic relation. It is important to note
however the high values, close to $1.0$ of these coefficients for the LTGs, which indicate the already established, strong correlation between the luminosity of the bulge / spheroidal component of 
a galaxy and $n_{b}$. The coefficents for ETGs are smaller but still indicate a strong correlation. Excluding the outliers has the result of increasing these 
coefficents to values similar to the ones corresponding to LTGs - $0.88$ and $0.89$ respectively.\\
\begin{table}
\begin{center}
\caption{\label{tab:black_holes} Black-hole masses for galaxies in our sample, as compiled from the literature: $a$ - \protect\cite{Savo15}, $b$ - \protect\cite{Bei12},
$c$ - \protect\cite{Gra13} (predicted), $d$ - \protect\cite{Do14} (estimated), $e$ - \protect\cite{Cara10} catalog (estimated); $f$ - \protect\cite{Bosch16}; 
$g$ - \protect\cite{Dullo20} (predicted). In \protect\cite{Bei12} and \protect\cite{Do14} there were no estimated/derived uncertainties for the black hole masses, while in 
\protect\cite{Dullo20}, an error of 0.85 dex on $\log(M_{BH})$ was considered for all their estimated masses. \protect\cite{Cara10} have a rough uncertainty of 50\protect\% 
for all their derived masses.} 
\begin{tabular}{{r|r|r}}
 \hline \hline
 $Galaxy$ &  $M_{BH}(x10^{8}M_{\odot})$  &   $\sigma_{M_{BH}}(x10^{8}M_{\odot})$ \\
          &    $[M_{\odot}]$                        &    $[M_{\odot}]$        \\
 \hline
NGC3031   &   $ 0.74^{a}  $   &       0.21  	 \\
NGC4594   &   $ 6.40^{a}  $   &       0.40     \\
NGC4736   &   $ 0.06^{b}  $   &       0.014 	 \\
NGC4826   &   $ 0.016^{a} $   &       0.004 	 \\
NGC0628   &   $ 4.90^{c}  $   &       1.00	 \\       
NGC3938   &   $ 5.00^{c}  $   &       1.00     \\
NGC2841   &   $ 1.10^{d}  $   &       0.00     \\ 
NGC3190   &   $ 2.14^{e}  $   &       1.07     \\
NGC4450   &   $ 2.20^{b}  $   &       0.28    \\ 
NGC7331   &   $ 1.04^{f}  $   &       0.43    \\ 
NGC3621   &   $ 0.01^{f}  $   &       0.008    \\ 
NGC5055   &   $ 8.32^{f}  $   &       1.91    \\ 
NGC4254   &   $ 0.097^{g} $   &       0.85    \\
NGC5033   &   $ 1.047^{g} $   &       0.85    \\
NGC5474   &   $ 0.0022^{g}$   &       0.85    \\
NGC7793   &   $ 0.0046^{g}$   &       0.85    \\
NGC0024   &   $ 0.026^{g} $   &       0.85    \\
NGC2976   &   $ 0.126^{g} $   &       0.85     \\
NGC0584   &   $ 2.60^{e}  $   &       0.00     \\
NGC0855   &   $ 0.042^{e} $   &       0.021   \\
NGC1404   &   $ 1.97^{d}  $   &       0.98    \\
NGC3265   &   $ 0.083^{e} $   &       0.041   \\
NGC4125   &   $ 3.95^{e}  $   &       1.97     \\
NGC4552   &   $ 4.70^{a}  $   &       0.50 	 \\
NGC1377   &   $ 0.29^{e}  $   &       0.14    \\
NGC1482	  &   $ 1.06^{e}  $   &       0.53    \\
NGC1705   &   $ 0.006^{e} $   &       0.003	\\
NGC3773   &   $ 0.036^{e} $   &       0.018    \\
NGC5866   &   $ 1.33^{e}  $   &       0.66   \\
 \hline
\end{tabular}
\end{center}
\end{table}
To check if this correlation is significant and if the two different slope relations found for the two samples of LTG and ETG are in fact statistically different,
we perform two statistical tests. First, we calculate a t-value based on the correlation coefficients previously derived for this relation, using the formula:
\begin{eqnarray}\label{eq:t_rcorr_value}
t(r^{corr})=r^{corr}\sqrt{n-2}/\sqrt{1-(r^{corr})^2}
\end{eqnarray}
where $n$ is the number of galaxies (early-type or late-type) and $r^{corr}$ is the Pearson correlation coefficient. We do this for both observed and intrinsic relations. The values are 
displayed in Table~\ref{tab:BH_scal_rel_statistics}, together with their associated $p$-values. As one can see, in all cases, the $p$-values for this relation are lower or much lower
than the standard $p=0.05$ value (alternatively $95\%$ confidence level), which is generally accepted as being a reliable indicator of the statistic significance of a result - in this case, 
of the correlation. Then, working in the null hypothesis that the two samples of galaxies have the same mean and should be described by one single $M_{b}-n_{b}$ correlation, we perform an
independent two sample t-test to calculate the t-statistics (or t score), for the pair of observed (for LTG and ETG) relations using all the data in the samples (including outliers) and then 
for the other pair, corresponding to the intrinsic relations. The t-values are also shown in Table~\ref{tab:BH_scal_rel_statistics}. Based on this, and taking into account the number of degrees 
of freedom for our sample, we then derive the associated $p$-values for both situations. The corresponding $p$-values (one-tail probability) are 0.085 for the pair of observed correlations and
0.064 for the intrinsic pair. A confidence level of at least $95\%$ (and thus p<0.05) would correspond to a t-statistics higher than 1.706. We can thus say that the two relations for ETG and
LTG are statistically different only at a $p=0.1$ significance level. If one repeats the test and excludes the extreme outliers in the calculation of t-score, this results in $p$-values of 0.025 
and 0.033, which means the two relations are statistically different at $95\%$ confidence level. Better statistics (a larger sample) would most likely produce an improvement in the statistical 
significance of this conclusion, that there are two different $M_{b}-n_{b}$ relations for LTG and ETGs.  

The second relation that we present and analyse in this section is the $M_{BH} - M_{b}$ relation (or alternatively $M_{BH} - L_{b}$). This is displayed in the central panel of 
Fig.~\ref{fig:BH_scal_rel}, the symbols and color scheme being the same as for the previous relation. One can notice from this plot that galaxies with more luminous bulges host
more massive black-holes, as expected, with the increasing trend being linear. This is valid for all the galaxies in our sample, both late and early-type ones. We derive the best-fit parameters
of this relation by linear regression fitting. Here, the three spiral galaxies in the upper middle part of the plot, way far-off from the general trend and with their black-hole mass estimated, together 
with the early-type galaxy with the lowest black-hole mass were excluded from the fit and from the calculation of the correlation coefficients as the values are deemed highly uncertain. 
While we plot $M_{BH}$ vs $M_{b}$, taking into account that $M_{b}=-2.5\times\log(L_{b})$, the best-fit parameters
in Table~\ref{tab:BH_scal_rel_parameters} correspond to the $M_{BH}-\log(L_{b})$ correlation.\\
It can be seen that the bulges of spiral galaxies reside on a slightly steeper relation than the ones of early-type galaxies. This result was also found by \cite{Savo16a} 
and \cite{Gra13}. In contrast, \cite{Bei12} found that there is no significant variation of this relation for different morphological types. Comparing the slope
of our best-fit relation for the ETGs, $0.91\pm0.16$ (observed) and $0.87\pm0.17$ (corrected), with other studies in the literature, we see that these are consistent
within errors with those in \cite{Savo16a}, who found $\alpha=1.03\pm0.11$ on average, from their multiple regression methods. However, their relation for bulges of late-type galaxies, with $\alpha=2.88\pm0.68$,
is considerably steeper than either our observed or intrinsic relation - $1.32\pm0.15$ \& $1.00\pm0.23$. Considering that their analysed images are at $3.6\mu$m, where dust influence is not
significant, the diffence in the slopes cannot be attributed solely to dust and inclination effects. Possible explanations are discussed in the next section. \cite{Dullo20}
fitted a single relation for their whole sample of ETGs and LTGs, considering as well the bulge luminosity at $3.6\mu$m, and found a slope of $1.23\pm0.15$, consistent within 
errors with our best-fit parameters. In another study, \cite{Bei12} found considerably shallower slopes for their samples of galaxies (early and late types) than our results.\\
\cite{Zhu21} determined a similar slope value for the relation between $M_{BH}$ and the stellar mass of classical bulges and of the elliptical galaxies in their entirety 
(they consider ellipticals as multi-component galaxies, with a core and two extra components - extended envelopes). Taking into account only the core of the ellipticals, 
which the authors conclude that should be considered in the coevolution of black-holes and their elliptical hosts, they find a $M_{BH}-M_{core}$ relation with a slope of 
$1.22\pm0.11$ and a scatter of 0.41 dex. With a proper transformation of our $L_b$ into the stellar bulge masses, considering the mass-to-light ratio in B band, the result
of \cite{Zhu21} could be comparable and consistent within errors with our values for the slope and scatter for this relation. \cite{KHo13} excluded pseudo-bulges from the
the $M_{BH} - M_{b}$ relation because of their different formation scenario, and because the black hole mass correlates more tightly with the mass of classical bulges and 
elliptical galaxies, and not with the ones of pseudo-bulges and discs.\\
The changes in the zero-point and slope when correcting for all the aforementioned effects are much more significant for the late-type bulges than for the early-types, as expected, with the 
corrected magnitudes shifting to higher values, which decrease the slope of the best-fit relation. The change is much less significant in the case of early-type galaxies.\\
As in the case of the first relation, since the slopes for the two samples of late-type galaxies have close enough values, we check again the statistical significance of the $M_{BH}-L_{b}$ relations derived,
and if the two relations are statistically different from each other (or in other words, if the means of the two samples are different). We use the same formula from Eq.~\ref{eq:t_rcorr_value} and the t-score
to investigate this. The respective values are presented in the same table. One can see from the first two columns that this relation is also statistically significant, having the $p$-value associated with the
$t(r^{corr})$ value much lower than the 0.05 standard. Testing the second issue, we see that the pair of observed relations for late-type and early-type galaxies are in this case significantly different
only at $p<0.1$ level (p=0.086), while for the intrinsic pair, this is not the case, with the associated p-value being larger than 0.10. We can thus conclude that the two
relations are not statistically different at p<0.05 level. 
\begin{table}
\begin{center}
 \caption{\label{tab:BH_scal_rel_parameters} The linear regression best-fit parameters for the black-hole scaling relations, for the late type and early type galaxies shown in Fig.~\ref{fig:BH_scal_rel}, both the 
 measured (observed) and corrected (intrinsic) ones: $\alpha$ - the intercept, $\beta$ - the slope. We also present here in the last two columns, the scatter of the relations, $\sigma$, and Pearson correlation
 coefficient, $r^{corr}$. }
 \begin{tabular}{{r|r|r|r|r}}
  \hline \hline
 BH relation      	&  $\alpha$  &  $\beta$  &  $\sigma$ (rms)  &   $r^{corr}$\\
 \hline
 $L_{b} - n_{b}$ (obs., LTG)    &  $7.02\pm0.06$   & $1.97\pm0.16$ &  0.52 &  0.90  \\
 $L_{b} - n_{b}$ (intrin., LTG) &  $6.62\pm0.08$   & $2.36\pm0.18$ &  0.61 &  0.92  \\
 $L_{b} - n_{b}$ (obs., ETG)    &  $7.12\pm0.14$   & $2.44\pm0.39$ &  0.58 &  0.67  \\
 $L_{b} - n_{b}$ (intrin., ETG) &  $6.91\pm0.17$   & $2.68\pm0.45$ &  0.56 &  0.66  \\
 \hline
 $M_{BH} - L_{b}$ (obs., LTG)    &  $6.33\pm0.26$  & $1.32\pm0.15$ & 0.39  &  0.84  \\
 $M_{BH} - L_{b}$ (intrin., LTG) &  $7.01\pm0.33$  & $1.00\pm0.17$ & 0.53  &  0.80  \\
 $M_{BH} - L_{b}$ (obs., ETG)    &  $7.36\pm0.27$  & $0.91\pm0.16$ & 0.35  &  0.91  \\
 $M_{BH} - L_{b}$ (intrin., ETG) &  $7.33\pm0.25$  & $0.87\pm0.17$ & 0.38  &  0.90  \\
 \hline
 $M_{BH} - n_{b}$ (obs., LTG)    &  $7.32\pm0.09$  & $3.42\pm0.37$ & 0.49  &  0.80  \\
 $M_{BH} - n_{b}$ (intrin., LTG) &  $7.02\pm0.12$  & $3.08\pm0.35$ & 0.43  &  0.83  \\
 $M_{BH} - n_{b}$ (obs., ETG)    &  $6.89\pm0.19$  & $3.14\pm0.54$ & 0.35  &  0.81  \\
 $M_{BH} - n_{b}$ (intrin., ETG) &  $6.57\pm0.22$  & $3.79\pm0.59$ & 0.33  &  0.76  \\
  \hline               
 \end{tabular}
 \end{center}
\end{table}

\begin{table}
\begin{center}
 \textbf{\caption{\label{tab:BH_scal_rel_statistics} The statistical tests parameteres for the black-hole scaling relations, for the late type and early type galaxies, corresponding to 
 Table~\ref{tab:BH_scal_rel_parameters}, for both the measured (observed) and corrected (intrinsic) ones: columns 1-2 -  $t(r^{corr})$-test value, derived from the formula
 presented in Eq.~\ref{eq:t_rcorr_value} and the corresponding \textit{p}-value; columns 3-4 - the t-test score (independent two-sample t-test) and its corresponding \textit{p}-value.}}
 \begin{tabular}{{r|r|r|r|r}}
  \hline \hline
 BH relation    &  $t(r^{corr})$  &  $p$-value  &  $t$-test  &   $p$-value   \\
 \hline
 $L_{b} - n_{b}$ (obs., LTG)    & 7.84   & <<0.05  &  1.41 & 0.085  \\
 $L_{b} - n_{b}$ (intrin., LTG) & 8.83   & <<0.05  &  1.57 & 0.064  \\
 $L_{b} - n_{b}$ (obs., ETG)    & 2.73   &  0.011  &  1.41 & 0.085  \\
 $L_{b} - n_{b}$ (intrin., ETG) & 2.63   &  0.013  &  1.57 & 0.064  \\
 \hline
 $M_{BH} - L_{b}$ (obs., LTG)    & 5.75  & <<0.05  & 1.40  &  0.086  \\
 $M_{BH} - L_{b}$ (intrin., LTG) & 1.91  & <<0.05  & 1.09  &  0.143  \\
 $M_{BH} - L_{b}$ (obs., ETG)    & 2.44  & <<0.05  & 1.40  &  0.086  \\
 $M_{BH} - L_{b}$ (intrin., ETG) & 2.61  & <<0.05  & 1.09  &  0.143  \\
 \hline
 $M_{BH} - n_{b}$ (obs., LTG)    & 5.08  & <<0.05  & 1.25  &  0.111  \\
 $M_{BH} - n_{b}$ (intrin., LTG) & 5.69  & <<0.05  & 1.27  &  0.107  \\
 $M_{BH} - n_{b}$ (obs., ETG)    & 4.09  &  0.004  & 1.25  &  0.111  \\
 $M_{BH} - n_{b}$ (intrin., ETG) & 3.48  &  0.009  & 1.27  &  0.107  \\
  \hline              
 \end{tabular}
 \end{center}
\end{table}
Coming back to the first relation that we mentioned in this section, we show it in the right-hand panel of Fig.~\ref{fig:BH_scal_rel}, plotted as $M_{BH}$ vs $\log(n_{b})$.
We have seen previously that early-type and late-type galaxies present two different slope relations in the $M_{b}-n_{b}$ relation and similar ones in the $M_{BH}-
M_{b}$ one. Here we check what is the case for this last relation. Previous recent work by \cite{Savo16b} showed that the values for the best-fit linear regression parameters for the two morphological
classes of galaxies are consistent with each other, within errors, meaning that only one relation exists for both LTGs and ETGs.\\
We overplot the best-fit relation with solid lines, in similar colors as in the other two panels, for both spirals and early-type galaxies. The best-fit linear regression
parameters are presented in Table~\ref{tab:BH_scal_rel_parameters}, as for the previous two relations. Here, a few points were excluded from the fit (namely the three
LTGs in the upper middle part of the plot and the two ETG in the opposite corners, far-off from the average trend), as the measurements of the black hole mass or S\'{e}rsic index
were unreliable and highly uncertain. This was done as before, to mitigate the effects of these outliers on the correlation coefficients or the best-fit parameters. At a
first look it is apparent that there are two similar, but slightly different trends between the black-hole mass and the S\'{e}rsic index of the bulge, for both the
observed and corrected case - a linear increase. To investigate this, we overplot the $\pm1\sigma$ limits just for the corrected relations (to avoid an unnecessary clutter
of the plot) in dashed lines. One can see a partial overlap of the two regions delimited by the standard deviations, especially at the higher-end of the relations. 
Keeping in mind the uncertainties of $\log(n_{b})$ from Sect.~\ref{sec:errors}, which are $\pm0.2$ dex at most (with the average being shown on the plot), and the ones 
associated with $M_{BH}$, but also the slopes of the two relations, we could draw the same conclusion as \cite{Savo16b}. However, considering the large scatter in the data
for late-type galaxies and the number of ETGs analysed, it is certain that more data would be beneficial for a more definitive conclusion.\\
Still, to investigate further this issue, and the statistical significance of this correlation, we perform similar statistical tests as in the case of the first two
black hole relations. The respective parameters can be found in the same Table~\ref{tab:BH_scal_rel_statistics}. One can see from the $t,p$ values that the relations are 
statistically significant. Then, working as before in the null hypothesis that the two relations are not different (and have the same mean), the low t-score and high
associated $p$-value leads us to the conclusion that at least at p<0.1 ($<90\%$ CL) level, the two relations are not different.\\
Here we would like to note the consistency (within errors) of our best-fit parameters for the slope and intercept of the relations with those derived by \cite{Savo16b} and
\cite{Sahu20}. Moreover, the scatter in the $M_{BH}$ direction for the last two relations is also comparable, though in some cases slightly larger, than the values found
by \cite{Savo16b} or \cite{Sahu20} for their larger sample of early and late-type galaxies. The Pearson coefficents indicate a lower degree of correlation, not 
as strong as for the $M_{b}-n_{b}$ relation. The scatter is also quite large here, but also in the previous relation. The main reason for this is that the black hole
masses in Table~\ref{tab:black_holes} are measured with a variety of techniques and instrumentation, some are predicted while others estimated, and thus the measurements 
are not heterogeneous and uniform in quality. The large scatter in this relation was also observed by \cite{Bei12}, who also note that this correlation is poor,
with a low level of significance and a poor predictor for $M_{BH}$. Still, our findings suggests that this correlation exists despite the large scatter, attributed mainly
to the uncertainties associated with the black-hole masses and in some cases, the bulge S\'{e}rsic indices.\\
Overall, the last two relations are important in predicting black-hole masses in large sample of galaxies, then deriving the black hole / supermassive black hole mass
functions, and further, the black hole mass density for early-type and late-type galaxies. Therefore, to clarrify the issue of having two or one single $M_{BH}-n_{b}$ / 
$M_{BH}-L_{b}$ relations is essential. However, all existent studies rely on small catalogues of black-hole masses, determined, as in the case of our sample, with 
different methods, and at various wavelengths, affected by the effects of dust attenuation. As shown here, accurate bulge parameters and correcting the relations for these 
effects is important, and can add information to the issue of having two separate relations or a single one for all morphologies. However, better statistics and more 
robust measurements are needed for a definitive answer.
\begin{figure}
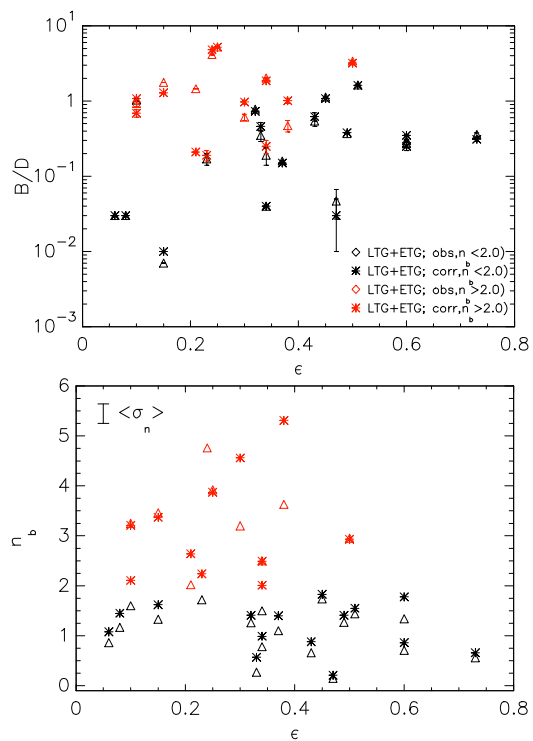

\begin{center}
 \includegraphics[scale=0.39]{Fig_6a.epsi}
 \hspace{-0.1cm}
 \includegraphics[scale=0.39]{Fig_6b.epsi}
 \caption{\label{fig:BD_nb_Qb} \textbf{\textit{Upper pannel}} The bulge-to-disc ratios of spirals and early-type galaxies vs. observed ellipticity of the bulges, $\epsilon$. 
 Black diamonds and stars represent the observed and intrinsic values for galaxies with $n_{b}<2.0$, while red diamonds and stars are for galaxies with $n_{b}\geq2.0$
 (see the overplotted legend). 
 The average uncertainty of $n_{b}$ is overplotted in the upper left corner, while the ones for ellipticity are not significant enough to be distinguished.
 The average uncertainty for $B/D$ is $0.025$. The error bars show the standard deviations, which are mostly of the size of the data points.
 \textit{Lower pannel} Bulge S\'{e}rsic indices vs. observed ellipticities of bulges. The symbols and color legend are similar as for the upper pannel plot.}
\end{center}
\end{figure}

\section{Discussion}\label{sec:discussion}

We are coming back in this section to a few issues observed when analysing the scaling relations presented in the previous section.

\subsection{Testing B/D vs ellipticity and $n_{b}$ vs ellipticity as viable bulge classification criteria}

We saw previously in the right-hand panel of Fig.~\ref{fig:Kormendy_relation} that using the Kormendy relation for the early-type galaxies in combination with a division
of our sample using a S\'{e}rsic index threshold did not produce a clear separation between spiral galaxies with classical and pseudo-bulges in the
$<\mu_{e,b}>-R_{e,b}$ plane. The main reason is that S\'{e}rsic index is not sufficient for this separation, apparently even when it is corrected for dust and inclination effects.
We test here a different approach to check if we can distinguish between classical and pseudo-bulges considering their different geometry (e.g. the intrinsic shape) as classical
bulges are rounder, spheroidal shape, while pseudo-bulges are flatter, disc-like structures. Thus, we plot in upper pannel of Fig.~\ref{fig:BD_nb_Qb} the $B/D$
ratios as a function of the observed (measured) ellipticity of the bulges, $\epsilon=1-Q_{b}$, while in the lower pannel, we plot $n_{b}$ vs. $\epsilon$. 
The bulge axis-ratio $Q_{b}$ is affected by projection effects, which are larger for bulges than for discs, as shown in \cite{Pas13a}. Still, we only plot
the observed values as we do not have neccesary corrections for this parameter. We note that if available, applying these corrections could produce significant changes, comparable
with those induced by projection effects on the bulge effective radius or S\'{e}rsic index. Our sample is again divided (in both plots) using the $n_{b}=2.0$ threshold 
in bulge S\'{e}rsic index. One can see in the upper plot that low $n_{b}$ galaxies show a large spread in ellipticity, from those characterstic to discs 
/ pseudo-bulges to lower ones, overlapping with the values characteristic for elliptical galaxies. Galaxies with high $n_{b}$ have ellipticity values similar with elliptical 
galaxies, with a smaller dispersion in values, occupying a separate locus in the $B/D-\epsilon$ plot. By analysing the intrinsic shape of bulges in a sample of nearby
galaxies, \cite{Cost18} found a similar result. Since low S\'{e}rsic index galaxies show a diversity in ellipticity values, and because of the relative overlap 
with the high $n_{b}$ bulges, we caution that using $B/D$ and $\epsilon$, with a threshold in $n_{b}$ to classify bulges will not produce an accurate, real division, as
among the $n_{b}<2.0$ bulges can hide galaxies with classical bulges. Looking at the lower plot in Fig.~\ref{fig:BD_nb_Qb}, one can see that low S\'{e}rsic index 
bulges show the whole range of ellipticities, as disc-like feature would present, while high $n_b$ bulges have lower ellipticities as classical bulges and elliptical bulges.
This is also in agreement with the studies of \cite{Cost17}, \cite{Cost18} and \cite{Gao20}. There is also a noticeable separation between the two populations of bulges in
the $n_{b}-\epsilon$ plot so one could conclude that this combination of parameters would be suitable to classify bulges. However, as mentioned earlier, we need to keep in
mind the projection effects on $Q_{b}$, which could bias this result. This is the reason why \cite{Gao20} cautioned that measured ellipticities of bulges should not be used
when trying to separate classical and pseudo-bulges.

\subsection{$n_{b}$-$B/T$ and $B/T$-$M_{\star}$ as proxies for bulge classification?}

We saw earlier in Fig.~\ref{fig:nb_vs_BT} and upper panel of Fig.~\ref{fig:BD_type_rel} from section \ref{sec:results} how the combinations $n_{b}$-$B/T$, and 
$B/T$-$M_{\star}$ can be used to morphologically separate galaxies into early- and late-type ones. One may ask if these criteria can be extended further, to be used to 
classify bulges, namely into classical and pseudo-bulges (disc-like). \\
It has been reported in the literature by \cite{Kor93}, \cite{Caro97} and \cite{Caro98} that most early-type galaxies, from S0 to Sab contain classical bulges, while pseudo-
bulges are very common starting with type Sb galaxies. Using the results for galaxy central components classification in HST (Hubble Space Telescope) V and H bands from 
\cite{Caro02} (which at that time had the best statistics), complemented with their classifications for the galaxies not fitted in \cite{Caro02} sample, \cite{KK04} find
the proportion of classical and pseudo-bulges to be as high as $50\%, 60\%$, and $44\%$ in S0-Sa, Sab and Sb galaxies. Starting with Hubble type Sb on, the overwhelming 
majority of late-type galaxies have a pseudo-bulge. \cite{KK04} also state in their criteria for recognising pseudo-bulges that bulges with small $B/T$ are not guaranteed 
to be disc-like bulges, but if $B/T$ is larger than $0.33\div0.5$ it can be concluded that the respective galaxy has a classical bulge, while a bulge with $B/T<1/3$ is 
likely to be a disc-like bulge. \cite{Gao20} find in their study that pseudo-bulges have in general a small bulge-to-total ratio (smaller than 0.1), while classical bulges
show a large spread in $B/T$, 26\% of their sample having also $B/T<0.1$. Because of the strong overlap at low $B/T$, the authors stress that bulge-to-total flux ratio 
alone is not a reliable parameter for separating bulges. A significant overlap in the distribution of bulge S\'{e}rsic indices for classical and pseudo-bulges is also found 
in the work of \cite{Gao20}. \cite{Gao18} and \cite{Gao20} report that up to $95\%$ of the bulges found in S0-Sab galaxies are classical, while the percentage of pseudo-
bulges is between $62\%$ for Sc galaxies and up to $90-100\%$ for late-type spirals.\\
Taking into account all this, and looking at Fig.~\ref{fig:nb_vs_BT}, imposing a combination of $n_{b}<2.0$ and $B/T<1/3$ selection criteria to distinguish the pseudo-bulges 
does separate most of the bulges found in the spiral galaxies of our sample, with two ETG bulges overlapping into this space. However, there are a few other bulges of late-type
galaxies which do not fit in this region. Imposing a more restrictive limit in bulge-to-total flux ratio (e.g. $B/T<0.1$, as previously mentioned studies found) produces
a less clean separation, now with more bulges of late-type galaxies outside the range and into the classical bulge zone. We appreciate this second combination of selection
criteria not to be accurate ($n_{b}<2.0$ \& $B/T<0.1$) in separating the distribution of pseudo-bulges in our sample. This could be due to the small size of our sample and
could potentially change in case of better statistics. \\
As \cite{Gao20} found that only ~10\% of their classified pseudo-bulges reside in galaxies with stellar masses higher than $10.5M_{\odot}$, if one uses a combination of 
selection thresholds of $B/T>1/3$ and $\log(M_{\star})>10.5$ in Fig.~\ref{fig:BD_type_rel} to separate classical bulges, an overlap with the more massive spirals in our sample is 
noticeable. Considering a less restrictive limit for $B/T$ (e.g. $B/T>0.1$) will produce more overlap between bulges in spirals and ETG. In any case, one notices that a
significant proportion of ETG bulges will be classified as pseudo-bulges, as the respective galaxy stellar masses are inferior to the imposed threshold. Again, a larger 
sample size and with better statistics would be beneficial to validate this combination of criteria to discriminate between bulges, but so as adding another threshold in
$n_{b}$ to the two selection thresholds.\\
Overall, detailed multi-component decomposition of galaxy images remain the most accurate method for determination of bulge types in galaxies. However, in case of large 
samples, when this process can be extremley time-consuming, and/or when the galaxy images do not have high-enough resolution and signal-to-noise ratio, proxies for bulge 
classification such as combination of parameters (e.g. $n_{b}-B/T$, with the associated thresholds) - could be used, albeit the uncertainties can be non-negligible. As 
pointed out by \cite{Neu17}, the more criteria are used together, the more accurate is the classification obtained.

\subsection{Potential sources of systematic errors and comparison with other studies}

Comparing with other similar studies, the small number of galaxies analysed here (especially the ETGs) could explain the larger scatter in the $M_{BH}-n_{b}$ or $M_{BH}-L_{b}$
relations and the differences in the best-fit parameters compared with the works of \cite{Savo13}, \cite{Savo16a}, \cite{Savo16b} or \cite{Sahu20}, for example. Another aspect is a different
method of calculation for the black hole masses in Table~\ref{tab:black_holes}, where some galaxies have directly measured $M_{BH}$, some are estimated while others are predicted,
with a high degree of uncertainty associated for the latter. This contributes as well to best-fit linear regression parameter values, their uncertainties and the scatter of the
respective relations.

Another aspect that could produce some differences in relation with other studies, is, as mentioned in Sect.~\ref{sec:fitting}, the consideration of ellipticals as
multicomponent objects, that contain a bulge and disc. While in previoulsy mentioned similar studies the ellipticals were treated as a single spheroidal component, practically
considering as galaxy only the cores of the elliptical galaxies, for our low-redhsift, high-resolution sample of elliptical galaxies, we consider our approach to be more
suitable and accurate. However, we do recognize that this can produce some differences in the results, compared with those in other works.

An important issue that may influence the parameters involved in the relations presented (black-hole relations included), and a potential source of systematic erorrs
is the fact that we only considered a bulge and a disc in our decomposition process, without any other extra-components, such as a nucleus, nuclear rings, 
lens/ovals, disk breaks or spiral arms. There are a number of previous works which underlined and shown the importance of inclusion of such components in the structural
analysis of galaxies, of course, if their presence is verified observationally, from an analysis of the residual images and/or surface brigthness or ellipticity profiles.\\
\cite{Lau09} revelead that lens/ovals are present in their analysed sample of early-type disc galaxies (S0-Sa) in large fractions, up to $86\pm6\%$ for unbarred lenticulars and
Sa0-Sa/Sab galaxies, with higher fractions for lenticulars. By doing multi-component decomposition on mid-infrared images of a medium sample of nearby barred spiral galaxies,
\cite{Kim14} included models with a disc break. They showed that in Type II disc galaxies (discs with a normal break/truncation in the outer parts), not including
a break can produce a derived disc scale-length with a intermediate value between the one of the inner and outer disc. At the same time, a part of the disc flux will be embedded
in the bulge or bar flux for the same type of disc galaxies, with $B/T$ suffering a $10\%$ decrease on average. On a similar note, \cite{Lau06} emphasized the importance of 
considering more components, especially in barred galaxies, otherwise the bulge mass and flux will be biased, with an overestimation of $B/T$. \\
\cite{Gao17} perform a detailed study of a small sample of representative, nearby, resolved disc galaxies, spanning all morphological types, with the purpose of investigating
the impact of choosing different combination of model functions for the various morphological features in galaxies, on the measured bulge parameters. The authors also 
quantify the errors induced by specific model choices and establish that outer lenses, outer rings and spiral arms can be safely ignored from a structural analysis, as
their effect on bulge parameters is at most $10\%$.  Nuclear and inner lenses/rings on the other hand, together with the disk breaks and bars, are found to have a more
significant effect on $B/T$ and $n_{b}$ and need to be considered. Nevertheless, the authors mention that is quite time-consuming to do such a thorough analysis on large 
samples of galaxies.

In light of this, we do recognize our simplified approach to the structural analysis of the galaxies in our sample. As mentioned in Paper I, we considered
only a two-component fit first because our method is suitable for this combination of fitting functions, while the dust and projection effects were quantified for the 
parameters of these components only. Another reason is that for the purpose of our study, we needed the structural and morphological parameters of bulges
as a whole, even though they might have some extra components embedded in them. Here we have only analysed unbarred galaxies (due to the lack of proper projection
and dust corrections for this component), and thus the systematic errors which arise when fitting only a bulge and a disk in a barred galaxy are not present in our results.
Taking into account the available classifications, galaxies with outer rings / lenses are not present in the analysed sample, although we cannot completely exclude them. In 
any case, as found by \cite{Gao17}, these features can be ignored. Thus the main source of systematic errors on bulge parameters in our analysis will be from not taking into
account the potential presence of disk breaks/truncations, nuclear and inner rings/lenses.\\
We conclude that our results for the galaxies which were analysed in similar studies might differ to a certain extent. However, this issue does not alter
significantly the main results and conclusions of this study.

\section{Summary and conclusions}\label{sec:conclusions}
 
We presented here a detailed analysis of the scaling relations of low-redshift bulges and early-type (ellipticals+lenticulars) galaxies and the biases introduced by the combined effects of dust,
inclination (projection) and decomposition on  the main parameters involved in these relations. Our study was performed on a representative sample of 29 spiral and
early-type galaxies (18 spirals and 11 ETGs) taken from the SINGS/KINGFISH survey, in B band. A thorough surface photometry and sky determination and subtraction was 
done to derive the integrated fluxes of each galaxy and its main components - discs and bulges. We derived the measured (observed) photometric and structural parameters of 
bulges (the spiral galaxies were analysed in Paper I) by doing a 2-component bulge-disc decomposition of galaxies, using GALFIT data analysis algorithm. Together with the
integrated fluxes and the structural parameters, we calculated the central, effective and mean effective bulge surface brightnesses, the absolute magnitudes and the bulge-to-disc ratios. 

The central face-on dust opacities in B band ($\tau_{B}^{f}$) were derived using the same method as in Paper I, for lenticular galaxies only. Subsequently, we have used the
numerical results for projection, dust and decomposition effects derived in \cite{Pas13a} and \cite{Pas13b} to correct all the necessary photometric and strucutural parameters
to obtain their intrinsic values. We then derived bulge and early-type scaling relations, such as the Kormendy relation, central bulge surface brightness - absolute magnitude, 
or the black-hole scaling relations, among others. Both the observed (measured) and the intrinsic (corrected) relations were presented, corrected for all the aforementioned effects,
in order to visualise the differences in the overall trend, their slopes and zero points. The scatter and the Pearson correlation coefficents were also determined, where possible,
to analyse the strength of the most import correlations, and a comparison with the values found in similar studies were presented. We performed statistical tests to investigate
the significance of the correlations, and if the relations found are statistically different for early- and late-type galaxies. By analysing these relations, our main 
conclusions are:
\begin{itemize}
 \item bulges of spiral galaxies reside on a steeper slope Kormendy relation that the early-type galaxies
 \item dust and inclination effects produce more important changes in the slope and zero-point of the Kormendy relation for spiral galaxies than those for ETGs
 \item if we divide our sample into galaxies with pseudo- or classical bulges on the basis of a S\'{e}rsic index threshold ($n_{b}=1.5$ or $2.0$), we cannot clearly notice
 pseudo-bulges residing far away (more than $3\sigma$) from the $<\mu_{e,b}>-R_{e,b}$ relation of the ETGs, as it was observed by other authors
 \item $n_{b}-B/T$ and $B/T-M_{\star}$ can be used when trying to classify galaxies into early and late types, as the two classes occupy different areas ($n_{b}-B/T$) or
 show different trends ($B/T-M_{\star}$); extending the use of these combinations of parameters (with specific thresholds) further to divide bulges into pseudo- or classical
 bulges, is problematic due to overlaps in the two distributions or large spread in values   
 \item we confirm two diferent intrinsic $M_{b}-n_{b}$ (or alternatively $L_{b}-n_{b}$) relations for late-type and early-type galaxies, with the latter showing a steeper 
 slope relation; the dust and inclination effects bias more the parameters of this relation (slope, intercept, scatter) for bulges of spiral galaxies than in the
 case of early-type ones;
 \item the black hole mass vs bulge luminosity - $M_{BH}-M_{b}$ (or $M_{BH}-L_{b}$) relation shows different best-fit parameters for bulges of LTGs and ETGs, with 
 the corresponding parameters for ETGs being consistent with other studies; while the slope is steeper for the bulges of late-type galaxies, the two relations are
 not statistically different at p<0.05 level, with only the observed relations are found to be different at p<0.1 probability;
 \item we can confirm within errors and taking into account the size of our sample, statistically similar $M_{BH}-n_{b}$ relations for both the early and late-type galaxies, 
 with more data needed for a decrease in the uncertainty of this result
 \item overall, our set of best-fit parameters (slope, zero-point, scatter) for the intrinsic set of black hole associated scaling relations (Fig.~\ref{fig:BH_scal_rel}) are consistent
 within errors with values found in other studies, while the Pearson coefficients derived indicate the strongest correlation for the $M_{b}-n_{b}$ relation,
 while the lowest degree of correlation was found for the $M_{BH}-n_{b}$ relation
\end{itemize}
The results presented here were obtained on a rather small sample of nearby spiral and early-type galaxies. Even so, we managed to draw some important conclusions and 
confirm a few of the previous results obtained from analysing larger samples. More data, especially for the black holes, will be beneficial for deriving more tight, less
scattered, intrinsic bulge scaling relations.

 \section*{Acknowledgements}
The author would like to thank the referee for a careful reading of the manuscript and for the useful suggestions which improved the quality and clarity of this paper.\\
This research made use of the NASA/IPAC Extragalactic Database (NED), which is operated by the Jet Propulsion Laboratory, California Institute of Technology, 
under contract with the National Aeronautics and Space Administration.\\
The author would like to acknowledge financial support from the Research and Inovation Ministery (MCI), project Nucleu- LAPLAS VI/2019.

\section*{Data Availability}
The data underlying this article are available in the article and in its online supplementary material.\\

\bsp	
\label{lastpage}


\begin{thebibliography}{}

\bibitem[Allen et al. (2006)]{All06} Allen, P. D., Driver, S. P., Graham, A. W., et al. 2006, MNRAS, 371, 2
\bibitem[Athanassoula (2005)]{Atha05} Athanassoula, E. 2005, MNRAS, 358, 1477
\bibitem[Beifiori et al. (2012)]{Bei12} Beifiori, A., Courteau, S., Corsini, E. M., Zhu, Y. 2012, MNRAS, 419, 2497
\bibitem[Bellstedt et al. (2017)]{Bell17} Bellstedt, S., Graham, A. W., Forbes, D. A., et al. 2017, MNRAS, 470, 1321
\bibitem[Bender (1990)]{Ben90} Bender, R. 1990, A\&A, 229, 441
\bibitem[Blakeslee et al. (2001)]{Bla01} Blakeslee, John P., Lucey, John R., Barris, Brian J., Hudson, Michael J., Tonry, John L. 2001, MNRAS, 327, 1004
\bibitem[Bournaud et al. (2007)]{Bou07} Bournaud, F., Elmegreen, B. G., \& Elmegreen, D. M. 2007, ApJ, 670, 237
\bibitem[Bournaud et al. (2009)]{Bou09} Bournaud, F., Elmegreen, B. G., \& Martig, M. 2009, ApJL, 707, L1
\bibitem[Bournaud (2016)]{Bou16} Bournaud, F. 2016, in "Galactic Bulges", ed. E. Laurikainen, R. Peletier \& D. Gadotti, 355
\bibitem[Brooks \& Christensen (2016)]{Brooks016} Brooks, A. \& Christensen, C. 2016, in "Galactic Bulges", ed. E. Laurikainen, R. Peletier \& D. Gadotti, 317
\bibitem[Capaccioli (1987)]{Capa87} Capaccioli, M. 1987, IAU Symp. 127, ``Structure and Dynamics of Elliptical Galaxies'', ed. P. T. de Zeeuw, 47
\bibitem[Caramete \& Biermann (2010)]{Cara10} Caramete, L. I. \& Biermann, P. L. 2010, A\&A, 521, A55
\bibitem[Carollo et al. (1997)]{Caro97} Carollo, C. M., Stiavelli, M., de Zeeuw, P. T., Mack, J. 1997, AJ, 114, 2366
\bibitem[Carollo et al. (1998)]{Caro98} Carollo, C. M., Stiavelli, M., Mack, J. 1998, AJ, 116, 68
\bibitem[Carollo et al. (2002)]{Caro02} Carollo, C. M., Stiavelli, M., Seigar, M., de Zeeuw, P. T., Dejonghe, H. 1998, AJ, 123, 159
\bibitem[Ciotti (1991)]{Cio91} Ciotti, L. 1991, A\&A, 249, 99
\bibitem[Ciotti \& Bertin (1999)]{Cio99} Ciotti, L. \& Bertin, G. 1999, A\&A, 352, 447
\bibitem[Costantin et al. (2017)]{Cost17} Costantin, L., Méndez-Abreu, J., Corsini, E. M. et al. 2017, A\&A, 601, A84
\bibitem[Costantin et al. (2018)]{Cost18} Costantin, L., Méndez-Abreu, J., Corsini, E. M. et al. 2018, A\&A, 609, A132
\bibitem[Djorgovski \& Davis (1987)]{Djo87} Djorgovski, S. \& Davis, M. 1987, ApJ, 313, 59 
\bibitem[Do et al. (2014)]{Do14} Do T., Wright S.A., Barth A.J. et al. 2014, AJ, 147, 93
\bibitem[Dong \& De Robertis (2006)]{Dong16} Dong X. Y., De Robertis M. M., 2006, AJ, 131, 1236
\bibitem[Dressler et al. (1987)]{Dre87} Dressler, A., Lynden-Bell, D., Burstein, D., et al. 1987, ApJ, 313, 42
\bibitem[Dullo et al. (2020)]{Dullo20} Dullo, B. T., Bouquin, A. Y. K., Gil de Paz, A., Knapen, J. H., Gorgas, J. 2020, ApJ, 898, 83
\bibitem[Elmegreen et al. (2008)]{Elme08} Elmegreen, B. G., Bournaud, F., Elmegreen, D. M. 2008, ApJ, 688, 67
\bibitem[Erwin et al. (2015)]{Erw15} Erwin, P., Saglia, R. P., Fabricius, M. et al. 2015, MNRAS, 446, 4039  
\bibitem[Faber \& Jackson (1976)]{Fab76} Faber, S. M. \& Jackson, R. E. 1976, ApJ, 204, 668
\bibitem[Faber (1977)]{Fab77} Faber, S. M. 1977, in "Evolution of Galaxies and Stellar Populations", ed. B. M. Tinsley \& R. B. Larson, 157
\bibitem[Ferrarese \& Merritt (2000)]{Fer00} Ferrarese, L., Merritt, D. 2000, ApJ, 539, L9
\bibitem[Fisher \& Drory (2008)]{Fish08} Fisher, D. B., Drory, N. 2008, AJ, 136, 773
\bibitem[Fisher \& Drory (2016)]{Fish16} Fisher, D. B., Drory, N. 2016, in "Galactic Bulges", ed. E. Laurikainen, R. Peletier \& D. Gadotti, 418, 41
\bibitem[Gadotti (2009)]{Gad09} Gadotti, D. 2009, MNRAS, 393, 1531
\bibitem[Gadotti et al. (2010)]{Gad10} Gadotti A. D., Baes M., Falony S. 2010, MNRAS, 403, 2053
\bibitem[Gao \& Ho (2017)]{Gao17} Gao, H., \& Ho, L. C. 2017, ApJ, 845, 114
\bibitem[Gao et al. (2018)]{Gao18} Gao, H., Ho, L. C., Barth, A. J., Li, Z.-Y. 2018, ApJ, 862, 100
\bibitem[Gao et al. (2019)]{Gao19} Gao, H., Ho, L. C., Barth, A. J., Li, Z.-Y. 2019, ApJS, 244, 34
\bibitem[Gao et al. (2020)]{Gao20} Gao, Hua, Ho, L. C., Barth, A. J., Li, Z.-Y. 2020, ApJS, 247, 20
\bibitem[Gebhardt et al. (2000)]{Geb00} Gebhardt, K., Bender, R., Bower, G., et al. 2000, ApJ, 539, L13
\bibitem[Gott (1977)]{Gott77} Gott, III, J. R. 1977, ARA\&A, 15, 235
\bibitem[Graham (2001)]{Gra01} Graham, A. W. 2001, AJ, 121, 820
\bibitem[Graham \& Guzman (2003)]{Guz03} Graham A.W., Guzm\'{a}n R. 2003, AJ, 125, 2936
\bibitem[Graham \& Driver (2005)]{Gra05} Graham, A.W. \& Driver, S.P. 2005, PASA, 22, 118
\bibitem[Graham \& Driver (2007)]{Gra07} Graham, A. W., \& Driver, S. P. 2007, ApJ, 655, 77
\bibitem[Graham (2013)]{Gra2013} Graham, A. W. 2013, in "Planets, Stars and Stellar Systems", ed. T.D.Oswalt \& W.C.Keel, Vol. 6, 91
\bibitem[Graham \& Scott (2013)]{Gra13} Graham, A. W. \& Scott, N. 2013, ApJ, 764, 151)
\bibitem[Graham \& Worley (2008)]{Gra08} Graham, A. W. \& Worley, C. C. 2008, MNRAS, 388, 1708
\bibitem[Grossi et al. (2015)]{Grossi15} Grossi, M., Hunt, L. K., Madden, S. C. et al. 2015, A\&A, 574, A126
\bibitem[Grootes et al. (2013)]{Gro13} Grootes, M., Tuffs, R.J., Popescu, C.C. et al. 2013, ApJ, 766, 59
\bibitem[H\"{a}ussler et al. (2007)]{Hau07} H\"{a}ussler, B., McIntosh, D. H., Barden, M. et al. 2007, ApJS, 172, 615
\bibitem[Huang et al. (2013a)]{Hua13a} Huang, S., Ho, L. C., Peng, C. Y., Li, Z.-Y., Barth, A. J. 2013a, ApJ, 766, 47
\bibitem[Huang et al. (2013a)]{Hua13b} Huang, S., Ho, L. C., Peng, C. Y., Li, Z.-Y., Barth, A. J. 2013b, ApJL, 768, L28
\bibitem[Huang et al. (2016)]{Hua16} Huang, S., Ho, L. C., Peng, C. Y., Li, Z.-Y., Barth, A. J. 2016, ApJ, 821, 114
\bibitem[Jerjen et al. (2000)]{Jer20} Jerjen, H., Binggeli, B. \& Freeman, K. C. 2000, AJ, 119, 593
\bibitem[Kennicutt et al. (2011)]{Ken11} Kennicutt, R. C., Calzetti, D., Aniano, G. et al. 2011, PASP, 123, 1347 
\bibitem[Kennicutt et al. (2003)]{Ken03} Kennicutt, R. C., Armus, L., Bendo, G. et al. 2003, PASP, 115, 928
\bibitem[Kim et al. (2014)]{Kim14} Kim, T., Gadotti, D. A., Sheth, K., et al. 2014, ApJ, 782, 6
\bibitem[Kormendy (1977)]{Kor77} Kormendy, J. 1977, ApJ, 218, 333 
\bibitem[Kormendy (1993)]{Kor93} Kormendy, J. 1993, IAU Symposium 153, 209 
\bibitem[Kormendy \& Richstone (1995)]{Kor95} Kormendy, J., \& Richstone, D. 1995, ARA\&A, 33, 581
\bibitem[Kormendy \& Gebhardt (2001)]{KorG01} Kormendy, J., \& Gebhardt, K. 2001, in 20th Texas Symposium on relativistic astrophysics, eds. J. C. Wheeler, \& H. Martel, AIPC, 586, 363
\bibitem[Kormendy \& Kennicutt (2004)]{KK04} Kormendy, J., Kennicutt, Jr., R. C. 2004, ARA\&A, 42, 603
\bibitem[Kormendy \& Ho (2013)]{KHo13} Kormendy, J., Ho, L. C. 2013, ARA\&A, 51, 511
\bibitem[Kormendy (2016)]{Kor16} Kormendy, J. 2016, ASSL, 418, 431
\bibitem[Laurikainen et al. (2005)]{Lau05} Laurikainen, E., Salo, H., Buta, R. 2005, MNRAS, 362, 1319
\bibitem[Laurikainen et al. (2006)]{Lau06} Laurikainen E., Salo H., Buta R. et al. 2006, AJ, 132, 2634
\bibitem[Laurikainen et al. (2007)]{Lau07} Laurikainen E., Salo H., Buta R., Knapen J., 2007, MNRAS, 381, 401
\bibitem[Laurikainen et al. (2009)]{Lau09} Laurikainen, E., Salo, H., Buta, R., Knapen, J. H. 2009, ApJL, 692, L34
\bibitem[Laurikainen et al. (2010)]{Lau10} Laurikainen, E., Salo, H., Buta, R., Knapen, J. H., Comer\'{o}n, S. 2010, MNRAS, 405, 1089
\bibitem[Magorrian et al. (1998)]{Mago98} Magorrian, J., Tremaine, S., Richstone, D., et al. 1998, AJ, 115, 2285
\bibitem[Marconi & Hunt (2003)]{Marco03} Marconi, A., \& Hunt, L. K. 2003, ApJL, 589, L21
\bibitem[M\'{e}ndez-Abreu et al. (2017)]{Men17} M\'{e}ndez-Abreu, J., Ruiz-Lara, T., Sánchez-Menguiano, L., et al. 2017, A\&A, 598, A32
\bibitem[M\"ollenhoff et al. (2006)]{Mol06} M\"ollenhoff, C., Popescu, C. C., Tuffs, R. J. 2006, A\&A, 456, 941
\bibitem[Neumann et al. (2017)]{Neu17} Neumann, J., Wisotzki, L., Choudhury, O. S., et al. 2017, A\&A, 604, A30
\bibitem[Nieto et al. (1991)]{Nie91} Nieto, J.-L., Bender, R., Arnaud, J., \& Surma, P. 1991, A\&A, 244, L25
\bibitem[Oh et al. (2017)]{Oh17} Oh, S., Greene, J. E., \& Lackner, C. N. 2017, ApJ, 836, 115
\bibitem[Pastrav (2020)]{Pas20} Pastrav, B. A. 2020, MNRAS, 493, 3580
\bibitem[Pastrav et al. (2013a)]{Pas13a} Pastrav, B. A., Popescu, C. C., Tuffs, R. J., Sansom, A. E., 2013a, A\&A, 553, A80
\bibitem[Pastrav et al. (2013b)]{Pas13b} Pastrav, B. A., Popescu, C. C., Tuffs, R. J., Sansom, A. E. 2013b, A\&A, 557, A137
\bibitem[Peng et al. (2002)]{Peng02} Peng, C. Y., Ho, L. C., Impey, C. D., Rix, H.-W. 2002, AJ, 124, 266
\bibitem[Peng et al. (2010)]{Peng10} Peng, C. Y., Ho, L. C., Impey, C. D., Rix, H.-W. 2010, AJ, 139, 2097
\bibitem[Planck Collaboration (2016)]{Planck} Planck Colaboration 2016, A\&A, 594, A13
\bibitem[Popescu et al. (2011)]{Pop11} Popescu, C. C., Tuffs, R. J., Dopita, M. A. et al. 2011, A\&A, 527, A109
\bibitem[R\'{e}my-Ruyer et al. (2015)]{Remy15} R\'{e}my-Ruyer, A., Madden, S. C., Galliano, F. et al. 2015, A\&A, 582, A121 
\bibitem[Renzini (1999)]{Ren99} Renzini, A. 1999, in "The Formation of Galactic Bulges", ed. C. M. Carollo, H. C. Ferguson \& R. F. G. Wyse, Cambridge Univ. Press, 9
\bibitem[Rix \& White (1990)]{Rix90} Rix, H.-W., White, S. D. M. 1990, ApJ, 362, 52
\bibitem[Rix \& White (1992)]{Rix92} Rix, H.-W., White, S. D. M. 1992, MNRAS, 254, 389
\bibitem[Rodriguez-Gomez et al. (2017)]{RG17} Rodriguez-Gomez, V., Sales, L. V., Genel, S. et al. 2017, MNRAS, 467, 3083
\bibitem[Sabbi et al. (2018)]{Sabbi18} Sabbi, E., Calzetti, D., Ubeda, L. et al. 2018, ApJS, 235, 23
\bibitem[Sachdeva et al. (2017)]{Sach17} Sachdeva, S., Saha, K., Singh, H. P. 2017, ApJ, 840, 79
\bibitem[Sachdeva et al. (2019)]{Sach19} Sachdeva, S., Gogoi, R., Saha, K., Kembhavi, A., Raychaudhury, S. 2019, MNRAS, 487, 1795
\bibitem[Sachdeva et al. (2020)]{Sach20} Sachdeva, S., Ho, L. C., Li, Y. A., Shankar, F. 2020, ApJ, 899, 89 
\bibitem[Sahu et al. (2020)]{Sahu20} Sahu, N., Graham, A. W., Davis, B. L. 2020, ApJ , 903, 97
\bibitem[Savorgnan et al. (2013)]{Savo13} Savorgnan, G., Graham, A. W., Marconi, A. et al. 2013, MNRAS, 434, 387
\bibitem[Savorgnan \& Graham (2015)]{Savo15} Savorgnan G.A.D., Graham A.W. 2015, MNRAS, 446, 2330
\bibitem[Savorgnan et al. (2016)]{Savo16a} Savorgnan, G., Graham, A. W., Marconi, A., Sani, E. 2016, ApJ, 817, 21
\bibitem[Savorgnan (2016)]{Savo16b} Savorgnan G.A.D. 2016, ApJ, 821, 88
\bibitem[Schlafly \& Finkbeiner (2011)]{SF11} Schlafly, E. F. \& Finkbeiner, D. P. 2011, ApJ, 737, 103
\bibitem[Schegel et al. (1998)]{Schl98} Schlegel D. J., Finkbeiner D. P., Davis M., 1998, ApJ, 500, 525
\bibitem[Simard et al. (2002)]{Sim02} Simard, L., Willmer, C. N. A., Vogt, N. P. et al. 2002, ApJS, 142, 1
\bibitem[Skibba et al. (2011)]{Ski11} Skibba, R. A., Engelbracht, C. W., Dale, D. . et al. 2011, ApJ, 738, 89 
\bibitem[Theureau et al. (2007)]{Theu07} Theureau, G., Hanski, M. O., Coudreau, N., Hallet, N., Martin, J. -M 2007, A\&A, 465, 71
\bibitem[Tonini et al. (2016)]{Ton16} Tonini, C., Mutch, S. J., Croton, D. J., Wyithe, J. S. B. 2016, MNRAS, 459, 4109
\bibitem[Tully \& Fisher (1988)]{Tul88} Tully, R. Brent \& Fisher, J. Richard 1988, "Catalog of Nearby Galaxies", Cambridge Univ. Press, ISBN 0521352991
\bibitem[Tully et al. (2013)]{Tul13} Tully, R. B., Courtois, H. M., Dolphin, A. E. et al. 2013, AJ, 146, 86
\bibitem[Tuffs et al. (2004)]{Tuf04} Tuffs, R. J., Popescu, C. C., V\"{o}lk, H. J., Kylafis, N. D., Dopita, M. A. 2004, A\&A, 419, 821
\bibitem[van den Bosch (2016)]{Bosch16} van den Bosch, Remco C. E 2016, ApJ, 831, 134
\bibitem[Weingartner \& Draine (2001)]{Wei01} Weingartner, J.C. \& Draine, B.T. 2001, ApJ, 548, 296
\bibitem[Weinzirl et al. (2009)]{Wein09} Weinzirl, T., Jogee, S., Khochfar, S., Burkert, A., \& Kormendy, J. 2009, ApJ, 696, 411
\bibitem[Wilson et al. (2013)]{Wilson13} Wilson, C. D., Cridland, A., Foyle, K. et al. 2013, ApJL 776, 30
\bibitem[Young \& Currie (1994)]{You94} Young, C. K. \& Currie, M. J. 1994, MNRAS, 268, L11
\bibitem[Zhu et al. (2021)]{Zhu21} Zhu, P., Ho, L. C. \& Gao, H. 2021, ApJ, 907, 6
\bibitem[Zibetti \& Groves (2011)]{Zib11} Zibetti, S. \& Groves, B. 2011, MNRAS, 417, 812

\end{thebibliography}
\end{document}